\documentclass[journal]{IEEEtran}

\ifCLASSINFOpdf

\else

\fi
\usepackage{soul}
\usepackage[utf8]{inputenc}
\usepackage{amsmath}
\usepackage{mathtools}
\usepackage{braket}
\usepackage{graphicx}
\usepackage{tikz}
\usepackage{quantikz}
\graphicspath{{./figures/}}
\usepackage{amsfonts}
\usepackage{comment}
\usepackage[ruled,vlined,linesnumbered]{algorithm2e}
\let\oldnl\nl% Store \nl in \oldnl
\newcommand{\nonl}{\renewcommand{\nl}{\let\nl\oldnl}}% Remove line number for one line

\usepackage{csquotes}
\usepackage{hhline}
\usepackage{amssymb}
\usepackage{listings}
\usepackage{color}
\definecolor{codegreen}{rgb}{0,0.6,0}
\definecolor{codegray}{rgb}{0.5,0.5,0.5}
\definecolor{codepurple}{rgb}{0.58,0,0.82}
\definecolor{backcolour}{rgb}{0.95,0.95,0.92}

\usepackage{subcaption}
\usepackage{titlesec}

\usepackage{dcolumn}
\usepackage{tabularx}
\usepackage{hyperref}
\hypersetup{
    colorlinks=true,
    linktoc=all,
    linkcolor=black,
    citecolor=magenta,
}
\usepackage{longtable}
\usepackage{braket}
\newcolumntype{C}{>{\centering\arraybackslash}X}
\SetAlFnt{\small}
\usepackage[font=small]{caption}
\begin{document}
% correct bad hyphenation here
\hyphenation{op-tical net-works semi-conduc-tor}

%\title{Quantum Machine Learning for Classification of Transportation System Tasks}
\title{QNN-QRL: Quantum Neural Network Integrated with Quantum Reinforcement Learning for Quantum Key Distribution}
%\title{QNN\_AIoT: A Quantum Neural Network Based Augmented Intelligence of Things for Vehicle Road Cooperation Systems}
%QDNF\_AIoT: A Quantum Deep Neuro-Fuzzy Based Augmented Intelligence of Things for Vehicle Road Cooperation Systems
\author{Bikash~K.~Behera\thanks{B.~K. Behera is with the Bikash's Quantum (OPC) Pvt. Ltd., Mohanpur, WB, 741246 India, e-mail: (bikas.riki@gmail.com).}, Saif Al-Kuwari\thanks{Saif Al-Kuwari is with the Qatar Center for Quantum Computing, College of Science and Engineering, Hamad Bin Khalifa University, Doha, Qatar. e-mail: (smalkuwari@hbku.edu.qa).}, and Ahmed~Farouk\thanks{A.~Farouk is with the Department of Computer Science, Faculty of Computers and Artificial Intelligence, Hurghada University, Hurghada, Egypt. e-mail:(ahmed.farouk@sci.svu.edu.eg).}}%

%\markboth{\today}%
%{Azad \MakeLowercase{\textit{et al.}}: A Blockchain-based Quantum Binary Voting in Decentralized IoT}

\maketitle

\begin{abstract}
Quantum key distribution (QKD) has emerged as a critical component of secure communication in the quantum era, ensuring information-theoretic security. Despite its potential, there are practical issues in optimizing key generation rates, enhancing security, and incorporating QKD into deployable implementations. In this paper, we introduce a unique framework for incorporating quantum machine learning (QML) algorithms, notably quantum reinforcement learning (QRL) and quantum neural networks (QNN), into QKD protocols to improve key generation performance. We evaluate our proposed algorithms using standard evaluation metrics, including accuracy, precision, recall, F1 score, confusion matrices, and ROC curves. Our results demonstrate considerable improvements in key generation quality. Furthermore, the existing and proposed models are investigated in the presence of different noisy channels to evaluate their robustness. The proposed integration of QML algorithms into QKD protocols and their noisy analysis creates a new paradigm for efficient key generation, which advances the practical implementation of QKD systems.
\end{abstract}


\IEEEpeerreviewmaketitle

\section{Introduction}\label{QVP:Sec1}
Quantum key distribution (QKD) is a promising technology to realize secure communication in future networks. %It has been used in the application of IoT-enabled medical devices to decrease power consumption \cite{Das_EC_2024}. Device-independent quantum key distribution (DI-QKD) offers the ultimate standard for secure key exchange that mitigates many quantum hacking risks that threaten non-DI QKD systems \cite{Zapatero_QI_2023}. This systematically investigates security loopholes in commercial QKD devices, offering insights into current weaknesses and initiatives to strengthen secure key exchange over insecure channels \cite{Vicario_OC_2024}. This review discusses the use of QKD for secure key exchange, focusing on the BB84 protocol's simulation, performance improvements, and its application potentials \cite{Jasim_PCS_2015}. This survey presents the key challenges such as key rate, distance, cost, and practical security issues for commercial adoption of QKD and current solutions to them \cite{Diamanti_nQI_2016}. Quantum cryptography uses quantum mechanics for secure cryptographic tasks beyond QKD, such as quantum money \cite{Aaronson_CACM_2012}, quantum cheque \cite{Behera_QIP_2017}, randomness generations \cite{Ma_nQI_2016}, and secure and delegated computations \cite{Broadbent_DCC_2016}.
\begin{figure*}
    \centering
    \includegraphics[width=\linewidth]{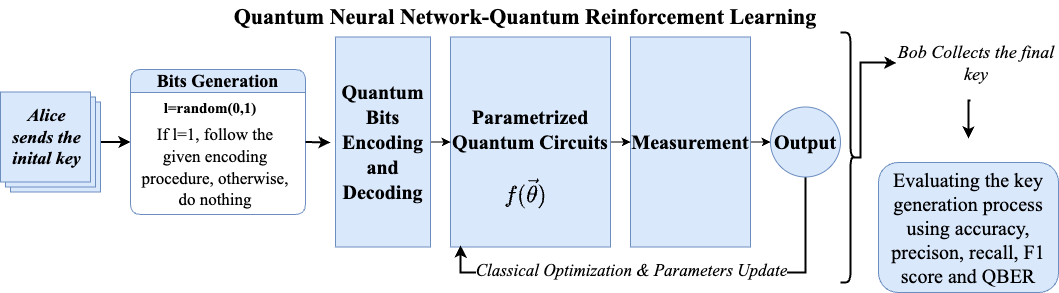}
    \caption{Schematic diagram showing the workings of the proposed QNN-QRL algorithm.}
    \label{fig:1}
\end{figure*}
However, QKD faces numerous challenges, including limited key generation rates, restricted communication distances, practical implementation complexities, and security vulnerabilities arising from imperfect devices, all of which hinder its widespread adoption in quantum-safe communications \cite{Lovic_CJSP_2020}. Key QKD protocols, such as BB84, B92, SARG04, and E91, have been developed to address some of these concerns \cite{Hong_In_2016}, but the practical use of QKD remains largely speculative. %Despite the potential of one-time pad encoding for secure communication, global scalability remains a challenge, as demonstrated by advances in metropolitan, backbone optical fiber, and satellite-based free-space QKD networks \cite{Zhang_OE_2018}. Efforts to miniaturize continuous-variable systems for practical applications aim to overcome limitations in operational distance and key generation rates \cite{Diamanti_ECOC_2019}. Rigorous metrology and robust classical authentication are crucial to mitigate vulnerabilities like side-channel attacks and ensure information-theoretic security over insecure channels \cite{Gui_JPCS_2022, Fregona_NATO}. Current implementations often rely on computational security, necessitating integration with post-quantum cryptography for practical applications \cite{Lella_Cr_2023}. Addressing routing complexities, dependency on trusted nodes, and scalability issues are vital for extending secure QKD communication over longer distances \cite{Kong_IEEECM_2024}. Future research focuses on improving speed, range, and resilience while ensuring seamless integration into existing infrastructures and effective management of complexity and key consumption in real-world deployments \cite{Sharma_CSDF_2024}.

Quantum machine learning (QML) approaches, such as quantum reinforcement learning (QRL) and quantum neural network (QNN) learning algorithms, offer promising solutions to the problem of QKD. QRL can improve resource allocation and routing in QKD networks, enabling efficient key generation and secure communication over long distances while decreasing vulnerabilities associated with trusted relay nodes \cite{Tsai_AS_2021, Kong_IEEECM_2024}. QNNs can detect and prevent side-channel attacks and device faults by learning intricate patterns in noisy quantum signals and forecasting probable vulnerabilities \cite{Gui_JPCS_2022, Sharma_CSDF_2024}. QRL-based frameworks can dynamically react to ambient conditions and device failures, improving scalability and robustness \cite{Lella_Cr_2023}. Integrating QNNs into authentication systems for conventional post-processing can improve resilience against denial-of-service attacks and reduce key consumption \cite{Fregona_NATO}. QML methods use quantum computing for real-time optimization and prediction, overcoming QKD's limitations in speed, scalability, and integration with existing networks \cite{Diamanti_ECOC_2019}. These enhancements demonstrate QML's transformative potential in overcoming QKD restrictions and paving the way for quantum-safe communication systems.

\subsection{Related Work}
This work integrates machine learning (ML) with a cascade error correction protocol that efficiently predicts quantum bit error correction and final key length \cite{qt_Al-Mohammed}. 

This uses deep learning models to detect and mitigate eavesdropping attacks \cite{Bommi_IEEE}, and reinforcement learning (RL) to adaptively alter system settings in real-time to account for the noise profiles of different quantum channels. This introduces QKD networks enabled by software-defined networking (SDN) as a promising technology \cite{Wang_MDPI_2019}. This explores the integration of IoT-enabled medical devices with QKD to enhance data security and energy efficiency in healthcare systems, particularly during pandemics like COVID-19 \cite{Das_EC_2024}. Three QKD protocols are demonstrated using IBM Q platforms \cite{Warke_QIP}. This reviews recent developments, in-field deployments, and standardization advancements of QKD networks, highlighting their potential and challenges for large-scale commercialization \cite{Stanley_QKD}. This introduces the principles and requirements of QKD in practical scenarios and applications while highlighting advancements in its technical and architectural domains \cite{Christoph_JP}.

This shows that RL can efficiently allocate resources in QKD networks \cite{Zuo_IEEE}.
This study investigates QKD as a use case for QML, enhancing eavesdropping attacks on the BB84 protocol, developing a unique cloning mechanism for noisy environments, and building a collective attack utilizing QKD post-processing \cite{Decker_arXiv}.
This study uses a Markov decision process and deep reinforcement learning (DRL) to optimize memory cutoff strategies for entanglement distribution over multi-segment repeater chains \cite{Simon_PRA_2023}. This proposes a DRL-based routing and resource assignment (RRA) scheme for QKD-secured optical networks (QKD-ONs), which outperforms traditional methods like deep Q-network (DQN), first-fit (FF), and random-fit (RF) across national science foundation network (NSFNET) and US backbone network (UBN24) \cite{Sharma_IETQC_2023}.
This framework combines QKD, Multi-Objective Optimization (MOO), and RL to ensure safe and efficient data transmission in Smart Grid Advanced Metering Infrastructure (SGAMI) \cite{Beula_OQE_2024}.
This demonstrates how RL optimizes resource allocation for efficient key distribution in QKD networks \cite{Zuo_ACPC_2020}.

\subsection{Contribution}
Fig. \ref{fig:1} represents the schematic of the overall proposed quantum algorithms for the quantum key generation process. Although there has been considerable previous work on improving the key generation process, QML algorithms such as QRL and QNN have not been integrated into them to investigate the quality of the key generation process. In this paper, we proposed two QRL-based QKD protocols and two QNN-based QKD protocols. The existing and proposed algorithms are run against six different noise channels to evaluate their performance. The contributions of the paper are summarized as follows.

\begin{enumerate}
    \item Two novel QRL-based algorithms are proposed for the quantum key generation process, namely QRL-V1 and QRL-V2.
    \item QNN-based enthronements are added to the popular BB84 and B92 protoocls are introuced to produce two novel algorithms ,  QNN-BB84, QNN-B92,
    \item Two integrated QNN-QRL algorithms are proposed, namely QNN-QRL-V.1, and QNN-QRL-V.2
    \item A comparative analysis between the existing BB84, B92, and the proposed QRL-based and QNN-based algorithms is performed
    \item An extensive noise model analysis for the existing and proposed algorithms is made. 
\end{enumerate}

\subsection{Organization}
The rest of the paper is organized as follows: Section \ref{background} briefly discusses the BB84, B92, QRL, and QNN algorithms. Section \ref{methodology} presents the general formulation of the problem and the proposed algorithms in detail. In Section \ref{results}, the settings, datasets, preprocessing, hyperparameters, and overall results are presented. Finally, the paper is concluded in Section \ref{conclusion}.

\section{Preliminaries}\label{background}
\subsection{BB84}
The BB84 protocol, a cornerstone of QKD leverages the principles of quantum mechanics to securely share a cryptographic key between two parties, traditionally called {Alice} (sender) and {Bob} (receiver), even in the presence of an eavesdropper, {Eve} \cite{Bennett_TCS_2014}.
The BB84 protocol exploits two fundamental properties of quantum mechanics, such as superposition and measurement, and the no-cloning theorem:
Alice selects a random sequence of bits, $b = \{b_1, b_2, \dots, b_n\}$, and a random sequence of bases, $r = \{r_1, r_2, \dots, r_n\}$, for encoding:
\begin{itemize}
    \item {Rectilinear basis} ($+$): $\ket{0}, \ket{1}$
    \item {Diagonal basis} ($\times$): $\ket{+} = \frac{1}{\sqrt{2}}(\ket{0} + \ket{1}), \; \ket{-} = \frac{1}{\sqrt{2}}(\ket{0} - \ket{1})$
\end{itemize}
For each bit $b_i$ and basis $r_i$:
\begin{itemize}
    \item If $r_i = +$: encode $b_i = 0$ as $\ket{0}$, $b_i = 1$ as $\ket{1}$.
    \item If $r_i = \times$: encode $b_i = 0$ as $\ket{+}$, $b_i = 1$ as $\ket{-}$.
\end{itemize}
Alice transmits the encoded qubits through a quantum channel to Bob.
Bob randomly selects bases $r' = \{r'_1, r'_2, \dots, r'_n\}$ to measure the incoming qubits:
\begin{itemize}
    \item If $r'_i = r_i$: Bob obtains the correct bit value $b_i$.
    \item If $r'_i \neq r_i$: Bob's measurement results are random.
\end{itemize}
Alice and Bob publicly announce their bases $r$ and $r'$ (but not the bit values $b$) over a classical channel. They discard measurements where $r_i \neq r'_i$.
The rest of the bits shared between Alice and Bob form the {sifted key}.
Alice and Bob reveal a subset of their sifted key to estimate the {quantum bit error rate (QBER)}:
\begin{eqnarray}
\text{QBER} = \frac{\text{Number of mismatched bits}}{\text{Total number of checked bits}}
\end{eqnarray}
If QBER exceeds a predefined threshold (e.g., $11\%$), they abort the protocol, suspecting eavesdropping.
Alice and Bob apply error correction and privacy amplification to derive the final shared key.
If Bob measures in the same basis as Alice:
\begin{eqnarray}
P(\text{correct}) = 1
\end{eqnarray}
If Bob measures in a different basis:
\begin{eqnarray}
P(\text{correct}) = \frac{1}{2}, \quad P(\text{incorrect}) = \frac{1}{2}
\end{eqnarray}
%\subsubsection{Impact of Eavesdropping}
If Eve intercepts and measures the qubits, she introduces errors due to the no-cloning theorem:
\begin{eqnarray}
\text{Probability of introducing an error per qubit: } \frac{1}{4}.
\end{eqnarray}
\subsection{B92}
The B92 protocol, proposed by Bennett in 1992, is a simplified version of the BB84 QKD protocol \cite{Bennett_PRL_1992}. Unlike BB84, which uses two orthogonal bases, the B92 protocol uses two non-orthogonal quantum states to encode a secret key. This simplification reduces the complexity of the protocol while retaining the security advantages of quantum mechanics.
The B92 protocol leverages two primary quantum mechanical principles, such as non-orthogonal states and the no-cloning theorem:
Alice chooses a random binary sequence $b = \{b_1, b_2, \dots, b_n\}$, where $b_i \in \{0, 1\}$. She encodes these bits using two non-orthogonal states:
\begin{itemize}
    \item $b_i = 0$: $\ket{\psi_0} = \ket{0}$
    \item $b_i = 1$: $\ket{\psi_1} = \ket{+} = \frac{1}{\sqrt{2}}(\ket{0} + \ket{1})$
\end{itemize}
These states are chosen such that:
\begin{eqnarray}
\braket{\psi_0 | \psi_1} = \frac{1}{\sqrt{2}},
\end{eqnarray}
indicating that they are not orthogonal.
Alice sends the encoded quantum states $\ket{\psi_i}$ to Bob over a quantum channel, who randomly measures each received qubit in one of two bases:
\begin{itemize}
    \item Rectilinear basis ($+$): $\{\ket{0}, \ket{1}\}$
    \item Diagonal basis ($\times$): $\{\ket{+}, \ket{-}\}$
\end{itemize}
If Bob's measurement yields a conclusive result (i.e., $\ket{0}$ or $\ket{-}$), he assigns a corresponding bit value:
\begin{itemize}
    \item Measurement outcome $\ket{0}$ corresponds to $b = 0$.
    \item Measurement outcome $\ket{-}$ corresponds to $b = 1$.
\end{itemize}
If Bob's measurement is inconclusive, he discards the result.
Bob communicates with Alice over a classical channel to confirm which measurements were conclusive. They discard bits corresponding to inconclusive measurements. 
Alice and Bob reveal a subset of their sifted key to estimate the QBER.
If the QBER is within an acceptable range, they apply error correction and privacy amplification to generate the final secret key.
In the B92 protocol, the overlap between the two non-orthogonal states determines the security of the protocol:
\begin{eqnarray}
\braket{\psi_0 | \psi_1} = \frac{1}{\sqrt{2}},
\end{eqnarray}
implying that the states cannot be perfectly distinguished.
If Bob measures in the wrong basis, the probability of obtaining an inconclusive result is non-zero:
\begin{eqnarray}
P(\text{inconclusive}) = \left|\braket{\psi_0 | \psi_1}\right|^2 = \frac{1}{2}.
\end{eqnarray}
If Eve intercepts the qubits, she must guess the state, introducing errors with a probability proportional to the state overlap:
\begin{eqnarray}
P(\text{error introduced}) = 1 - \left|\braket{\psi_0 | \psi_1}\right|^2 = \frac{1}{2}.
\end{eqnarray}

\subsection{QRL}
QRL integrates quantum computing with RL to enhance decision-making and optimize learning tasks in dynamic environments. In classical RL, agents interact with environments, performing actions \(a_t\), observing states \(s_t\), and receiving rewards \(r_t\) to maximize cumulative returns \(G_t\):  
\begin{equation}
G_t = \sum_{k=0}^{\infty} \gamma^k r_{t+k},
\end{equation}  
where \( \gamma \) is the discount factor. The optimal policy \( \pi^*(s) \) maps states to actions, maximizing the expected return.  
QRL introduces quantum states \( |\psi_t\rangle \), represented as:  
\begin{equation}
|\psi_t\rangle = \sum_i c_i |i\rangle,
\end{equation}  
where \( |i\rangle \) and \( c_i \) are computational basis states and complex coefficients, respectively. Actions correspond to quantum operations, and rewards are encoded as quantum observables \( \hat{R} \), with the goal of maximizing:  
\begin{equation}
\langle \hat{R} \rangle = \langle \psi_t | \hat{R} | \psi_t \rangle.
\end{equation}  
The quantum value function \( V(s) \), encoding the expected cumulative reward, can be optimized using quantum algorithms, while the quantum Bellman equation relates value functions to expected rewards:  
\begin{equation}
V(s_t) = \mathbb{E} \left[ r_t + \gamma \langle \psi_{t+1} | \hat{H} | \psi_{t+1} \rangle \right],
\end{equation}  
where \( \hat{H} \) is the system's Hamiltonian at time \( t+1 \).

\subsection{QNN}
QNNs integrate quantum computing and machine learning, utilizing quantum mechanics to enhance neural network architectures. They leverage quantum properties such as superposition, entanglement, and interference to address problems infeasible for classical networks. This section outlines their structure, mathematical foundation, and advantages.
QNNs consist of quantum gates and circuits, similar to layers in classical neural networks. Input data \( x \) is encoded into a quantum state \( |x\rangle \) via an embedding circuit, processed through parameterized quantum gates, and measured to generate predictions.
The core operation is the parameterized quantum circuit \( U(x, \theta) \), expressed mathematically as:
\begin{eqnarray}
| \psi_{\text{output}} \rangle = U(x, \theta) | x \rangle,
\end{eqnarray}
where \( | \psi_{\text{output}} \rangle \) is the quantum state post-processing.
QNNs employ rotation gates and entangling gates. Single-qubit rotation gates are:
\begin{eqnarray}
R_x(\theta) = e^{-i \frac{\theta}{2} X}, R_y(\theta) = e^{-i \frac{\theta}{2} Y}, R_z(\theta) = e^{-i \frac{\theta}{2} Z},
\end{eqnarray}
with \( X, Y, Z \) being Pauli matrices and \(\theta\) the rotation angle.
Entangling gates like the controlled-NOT (\(\text{CNOT}\)) gate create quantum correlations:
\begin{eqnarray}
\text{CNOT} = |0\rangle\langle 0| \otimes I + |1\rangle\langle 1| \otimes X.
\end{eqnarray}

%\subsection{Training a QNN}
Training involves iteratively optimizing parameters \(\theta\) to minimize a loss function:
\begin{enumerate}
    \item Encode input \( x \) into \( |x\rangle \).
    \item Process \( U(x, \theta) \) to compute the quantum state.
    \item Measure the state for predictions \( y_{\text{predicted}} \).
    \item Compute the loss function:
    \begin{eqnarray}
    L(\theta) = \frac{1}{N} \sum_{i=1}^N \left( y_{\text{predicted}, i} - y_{\text{true}, i} \right)^2,
    \end{eqnarray}
    where \( N \) is the number of samples.
    \item Update \(\theta\) using gradient descent:
    \begin{eqnarray}
    \theta \leftarrow \theta - \eta \nabla_\theta L(\theta),
    \end{eqnarray}
    with \( \eta \) as the learning rate.
\end{enumerate}

QNNs offer unique benefits:
\begin{itemize}
    \item {Exponential Hilbert Space:} Efficiently process high-dimensional data.
    \item {Quantum Parallelism:} Evaluate multiple states simultaneously.
    \item {Entanglement:} Model complex feature correlations.
\end{itemize}
QNNs merge quantum computation with neural network design, creating a powerful tool for optimization, data science, and AI. With advancing quantum hardware, they hold the potential to solve challenging problems across various domains.

\section{Proposed Algorithms}\label{methodology}
%\subsection{Problem Overview\label{sec2}}

The objective is to develop a QRL-based QKD algorithm, integrate the QNN model into the existing and QRL-based algorithms, and propose novel algorithms that can outperform the existing algorithms for the QKD process. A key is generated between two parties, Alice and Bob, where they follow different strategies to achieve this task. Here, a key is taken as $\{k_1, k_2,...,k_n\}$, which is generated by Alice, and sent to Bob over a channel. Bob then performs some measurements on those keys and forms a set of received keys $\{l_1,l_2,...l_n\}$. Accuracy, precision, recall, and F1 score are calculated from these two sets of keys. Furthermore, QBER is calculated, which is the amount of unmatching bits. The goal of the algorithm is to maximize the evaluation metrics, whereas to minimize the QBER.

\begin{eqnarray}
\text{Evaluation metrics}=\text{max}\{\text{accuracy}, \text{precision}, \text{recall}, \text{F1 score}\}
\end{eqnarray}

\begin{eqnarray}
\text{Error rate}=\text{min}\{\text{QBER}\}
\end{eqnarray}

%\section{Methodology \label{sec3}}

\subsection{QRL Algoirthms}
%\subsection{QRL-V.1.: Quantum Reinforcement Learning (QRL) Key Agreement Algorithm}

This proposed QRL-enhanced key generation algorithm consists of the following steps:

\begin{itemize}
    \item[1)] \textbf{Quantum State Preparation by Alice:}  
    Alice randomly selects a classical bit (0 or 1) and encodes it into a quantum state using a Hadamard gate followed by a phase gate with a secret angle $\theta_1$. This operation transforms the computational basis states into superpositions with embedded phase information.
    
    \item[2)] \textbf{Reinforcement-Learning-Based Decoding by Bob:}  
    Bob receives the quantum state and applies a decoding strategy using a phase gate $P(-\theta_2)$ (with angle $\theta_2$ determined by the QRL algorithm) followed by a Hadamard gate. The decoding angle is iteratively learned through reinforcement learning to maximize the probability of correct measurement outcomes.
    
    \item[3)] \textbf{Learning Strategy (Q-Update and Reward):}  
    The agent observes the measurement outcomes based on the difference $\Delta \theta = \theta_1 - \theta_2$. The reward signal is computed using the measurement probabilities $P_0$ and $P_1$: a higher reward is assigned when the correct bit (as originally sent by Alice) is recovered with high confidence. Specifically, the reward function is defined as:
    \begin{equation}
        \text{Reward} = \max(P_0, P_1)
    \end{equation}
    The QRL agent updates its policy based on this reward, aiming to steer $\Delta \theta$ toward the optimal value (approximately 1.2 radians), where the distinguishability between bit values is maximized.
    
    \item[4)] \textbf{Final Key Extraction:}  
    Once convergence is achieved, i.e., the optimal $\theta_2$ is found such that the outcome probabilities clearly indicate Alice’s original bit, the bit is accepted into the final key. No classical communication is required in this process, preserving quantum communication advantages.
\end{itemize}

The transformations of quantum states throughout the protocol are as follows:

\begin{align}
    H\ket{0} &= \frac{1}{\sqrt{2}}(\ket{0} + \ket{1}) \\
    \xrightarrow[P(-\theta_2)]{P(\theta_1)} & \frac{1}{\sqrt{2}}(\ket{0} + e^{i\Delta \theta} \ket{1}) \\
    \xrightarrow[]{H} & \frac{1 + e^{i\Delta \theta}}{2} \ket{0} + \frac{1 - e^{i\Delta \theta}}{2} \ket{1}
\end{align}

The measurement probabilities are given by:

\begin{align}
    P_0 &= \frac{1 + \cos(\Delta \theta)}{2}, \quad 
    P_1 = \frac{1 - \cos(\Delta \theta)}{2}
\end{align}

\begin{figure}
    \centering
    \includegraphics[width=1\linewidth]{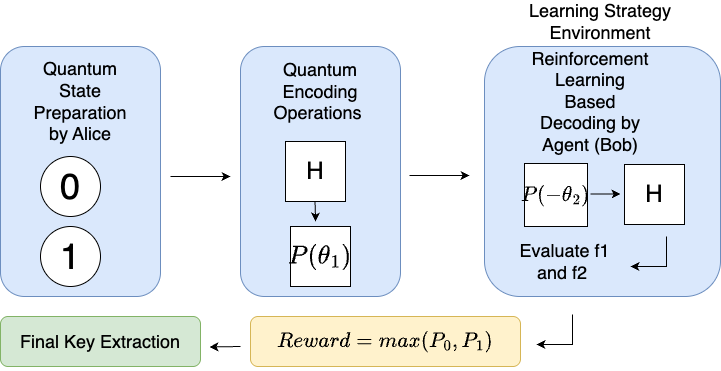}
    \caption{A schematic of QRL-V.1. algorithm.}
    \label{qrl-v1_schematic}
\end{figure}

\begin{figure}[]
\centering
\begin{subfigure}{\linewidth}
\includegraphics[width=\linewidth]{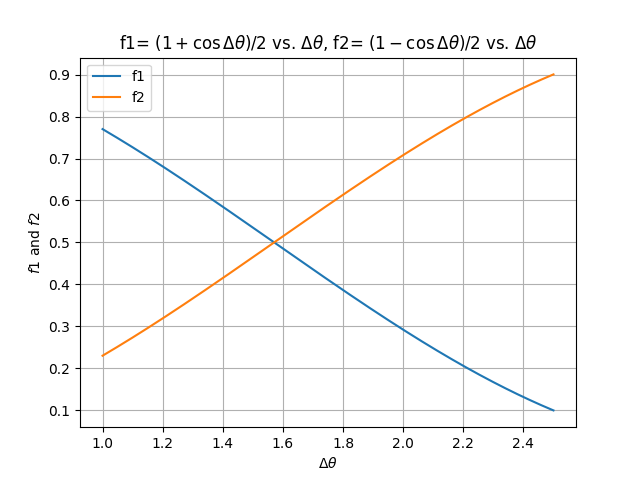}
    \caption{}
    \label{fig2a}
\end{subfigure}\hfill
\begin{subfigure}{\linewidth}
    \includegraphics[width=\linewidth]{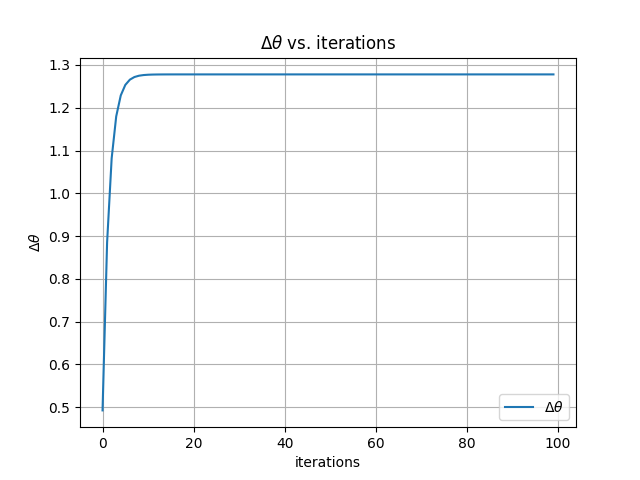}
\caption{}
\label{fig2b}
\end{subfigure}\hfill
\begin{subfigure}{\linewidth}
    \includegraphics[width=\linewidth]{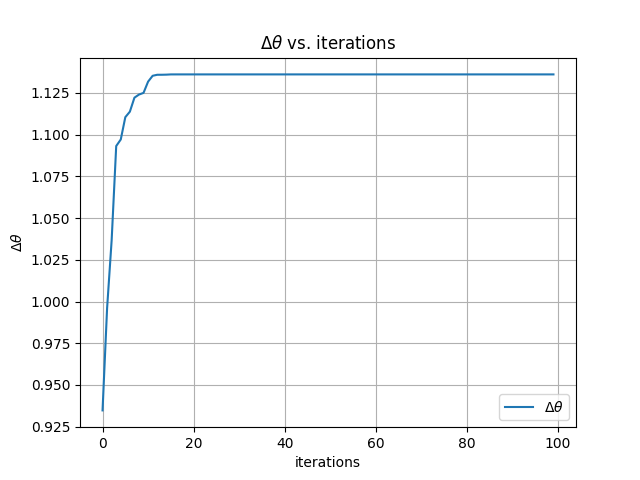}
\caption{}
\label{fig2c}
\end{subfigure}\hfill
\caption{(a) Function plots versus angle. $\Delta \theta$ vs. number of iterations for (b) QRL-V.1., and (c) QRL-V.2.}
\label{Fig2}
\end{figure}

\begin{figure*}[]
\centering
\begin{subfigure}{\linewidth}
\includegraphics[width=0.9\linewidth]{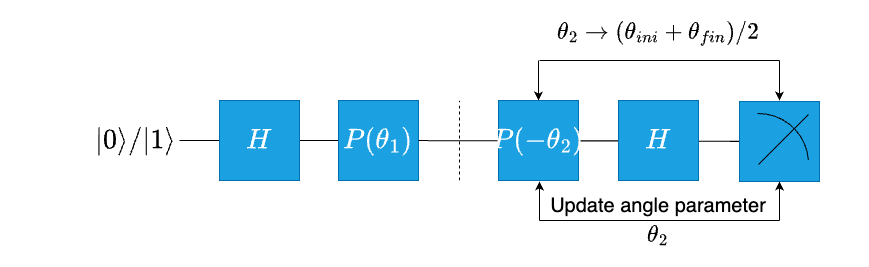}
    \caption{}
    \label{fig3a}
\end{subfigure}\hfill
\begin{subfigure}{\linewidth}
    \includegraphics[width=0.9\linewidth]{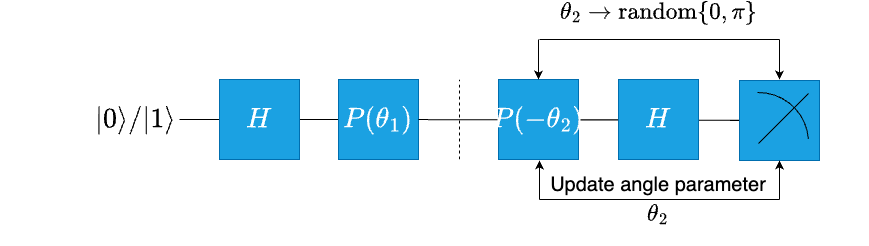}
\caption{}
\label{fig3b}
\end{subfigure}
\caption{Quantum circuits representing (a) QRL-V.1 and (b) QRL-V.2 algorithms.}
\label{Fig3}
\end{figure*}

The same derivation holds for the initial state $\ket{1}$. The QRL agent selects $\theta_2$ values that maximize $P_0$ when Alice sends $\ket{0}$ and $P_1$ when she sends $\ket{1}$. A binary search initialization followed by iterative refinement ensures convergence of $\Delta \theta$ to the optimal value (around 1.2), maximizing distinguishability. $P0_{\ket{0}}$, $P1_{\ket{0}}$, $P0_{\ket{1}}$, and $P1_{\ket{1}}$ are plotted against $\Delta \theta$, where it is found that when Alice sends 0, $P0_{\ket{0}} > P1_{\ket{0}}$ (let us say around the value of 1.2), and when Alice sends 1, $P0_{\ket{1}} < P1_{\ket{1}}$ around the same value. So, there is a clear distinction for extracting information about what Alice sends. The plots shown in Fig. \ref{fig2a} clearly represent the above strategy. The full quantum circuit is shown in Fig.~\ref{fig3a}. Learning convergence is illustrated in Fig.~\ref{fig2b}, where $\Delta \theta$ converges to the optimal range, confirming effective training and secure key inference without classical feedback. The overall schematic is shown in Fig. \ref{qrl-v1_schematic}.

\iffalse
\begin{algorithm}[!t]
\DontPrintSemicolon
\nonl \textbf{Input:} Number of key bits $N$\\
\nonl \textbf{Output:} Final key $F_k$

\For{$i = 1$ to $N$}{
  Sample a random bit $k_i \in \{0, 1\}$\\
  Sample a random angle $l_i \in [0, \pi]$
  
  Initialize search intervals: $n_1^i = 0$, $n_2^i = \pi/2$, $n_3^i = \pi$
  \For{$j = 1$ to $1000$}{
    \If{$k_i = 0$}{
      Prepare qubit in $|0\rangle$ state
    }\Else{
      Prepare qubit in $|1\rangle$ state
    }
    
    Apply Hadamard gate\\
    Apply phase rotation gate with angle $l_i$\\
    
    Calculate midpoints: $m_1^i = (n_1^i + n_2^i) / 2$, $m_2^i = (n_2^i + n_3^i) / 2$\\
    
    Apply controlled-Z gate with phase $m_1^i$\\
    Apply Hadamard gate\\
    
    Measure the qubit\\
    Update search intervals based on measurement outcome and angle differences\\
  }
  
  Prepare qubit based on $k_i$\\
  Apply Hadamard gate\\
  Apply phase rotation gate with angle $l_i$\\
  Apply controlled-Z gate with phase $(n_1^i + n_3^i) / 2$\\
  Apply Hadamard gate\\
  Measure the qubit\\
  Add the measurement result to $K_m$\\
}
%\State 
Compare $K_b$ and $K_m$ to form the final key $F_k$
\caption{Quantum Key Generation: QRL-V.1.}
\label{alg:qkg_rl}
\end{algorithm}  
\fi

\begin{algorithm}[!t]
\DontPrintSemicolon
\nonl \textbf{Input:} Number of key bits $N$\\
\nonl \textbf{Output:} Final secret key $F_k$

\For{$i = 1$ to $N$}{
  Alice randomly selects a bit $k_i \in \{0, 1\}$ \tcp*{Classical bit to encode}
  Alice selects a secret angle $\theta_1^i \in [0, \pi]$ \tcp*{Quantum encoding parameter}
  
  Bob initializes the search interval: $n_1^i = 0$, $n_3^i = \pi$, $n_2^i = \frac{n_1^i + n_3^i}{2}$\\
  Initialize Q-values: $Q(n_1^i) = Q(n_3^i) = Q(n_2^i) = 0$\\

  \For{$j = 1$ to MaxEpisodes}{
    \textbf{Environment:} Alice encodes $k_i$ using $H$ and $P(\theta_1^i)$ \\
    \textbf{Agent:} Bob applies $P(-\theta_2^i)$ with $\theta_2^i \in \{n_1^i, n_2^i, n_3^i\}$\\
    Bob applies Hadamard gate and performs measurement \\
    
    Compute $\Delta \theta^i = \theta_1^i - \theta_2^i$ \\
    Compute measurement probabilities: 
    $P_0 = \frac{1 + \cos(\Delta \theta^i)}{2}$, 
    $P_1 = \frac{1 - \cos(\Delta \theta^i)}{2}$\\
    
    \textbf{Reward} $\gets \max(P_0, P_1)$ \tcp*{Reward proportional to distinguishability}
    
    \textbf{Q-Update:} Update $Q(\theta_2^i) \leftarrow$ Reward \\
    Select next $\theta_2^i$ using binary search policy based on Q-values
    
    \If{convergence condition met}{
        \textbf{break}
    }
  }

  Final $\theta_2^i \gets \arg\max_{\theta} Q(\theta)$ \tcp*{Optimal decoding angle}
  Bob applies $P(-\theta_2^i)$ and Hadamard gate, measures qubit\\
  Add measured bit to key: $F_k \gets F_k \cup \text{measured\_bit}$
}

\caption{QRL-V.1: Quantum Reinforcement Learning Key Agreement Protocol}
\label{alg:qkg_rl}
\end{algorithm}

\iffalse
\begin{algorithm}[!t]
\DontPrintSemicolon
\nonl \textbf{Input:} Number of key bits $N$\\
\nonl \textbf{Output:} Final key $F_k$

\For{$i = 1$ to $N$}{
  Sample a random bit $k_i \in \{0, 1\}$\\
  Sample a random angle $l_i \in [0, \pi]$\\
  Initialize search intervals: $n_1^i = 0$, $n_2^i = \text{random}(0, \pi)$, $n_3^i = \pi$\\
  
  \For{$j = 1$ to $100$}{
    \If{$k_i = 0$}{
      Prepare qubit in $|0\rangle$ state
    }\Else{
      Prepare qubit in $|1\rangle$ state
    }
    
    Apply Hadamard gate\\
    Apply phase rotation gate with angle $l_i$\\
    
    Sample random midpoints: $m_1^i = \text{random}(0, n_2^i)$, $m_2^i = n_2^i + \text{random}(0, n_3^i - n_2^i)$\\
    
    Apply controlled-Z gate with phase $m_1^i$\\
    Apply Hadamard gate\\
    
    Measure the qubit\\
    Update search intervals based on measurement outcome and angle differences
  }
  
  Prepare qubit based on $k_i$\\
  Apply Hadamard gate\\
  Apply phase rotation gate with angle $l_i$\\
  Apply controlled-Z gate with phase $(n_1^i + n_3^i) / 2$\\
  Apply Hadamard gate\\
  Measure the qubit\\
  Add the measurement result to $K_m$
}
%\State 
Compare $K_b$ and $K_m$ to form the final key $F_k$
\caption{Quantum Key Generation: QRL-V.2.}
\label{alg:qkg_random_angles}
\end{algorithm}
\fi

\begin{algorithm}[!t]
\DontPrintSemicolon
\nonl \textbf{Input:} Number of key bits $N$\\
\nonl \textbf{Output:} Final key $F_k$

\For{$i = 1$ to $N$}{
  Alice selects a classical bit $k_i \in \{0, 1\}$\\
  Alice encodes using a secret phase angle $\theta_1^i \in [0, \pi]$\\
  
  Initialize search interval for Bob: $n_1^i = 0$, $n_3^i = \pi$, $n_2^i = \frac{n_1^i + n_3^i}{2}$\\
  Initialize reward table: $Q(n_1^i) = Q(n_2^i) = Q(n_3^i) = 0$
  
  \For{$j = 1$ to 100}{
    \textbf{Environment:} Alice prepares qubit in $\ket{k_i}$\\
    Apply Hadamard gate, then phase gate $P(\theta_1^i)$\\
    
    \textbf{Agent:} Bob samples $\theta_2^i \in \{n_1^i, n_2^i, n_3^i\}$\\
    Apply $P(-\theta_2^i)$ and Hadamard gate, then measure\\
    
    Compute $\Delta\theta^i = \theta_1^i - \theta_2^i$\\
    Compute $P_0 = \frac{1 + \cos(\Delta \theta^i)}{2}$, $P_1 = \frac{1 - \cos(\Delta \theta^i)}{2}$\\
    \textbf{Reward} $\gets \max(P_0, P_1)$\\
    
    \textbf{Q-Update:} $Q(\theta_2^i) \gets \text{Reward}$\\
    Update angle range $\{n_1^i, n_2^i, n_3^i\}$ by narrowing toward the maximum reward\\
    
    \If{convergence condition met}{
        \textbf{break}
    }
  }
  
  Set final angle $\theta_2^i = \arg\max_{\theta} Q(\theta)$\\
  Bob reconstructs the state using $P(-\theta_2^i)$, Hadamard gate, and measures\\
  Append the result to final key $F_k$
}
Compare $K_b$ and $K_m$ to finalize the key $F_k$

\caption{Quantum Key Generation: QRL-V.2 with Reinforcement-Based Angle Adaptation}
\label{alg:qkg_random_angles}
\end{algorithm}

\begin{figure*}[]
\centering
\begin{subfigure}{\linewidth}
\centering
    \includegraphics[width=0.9\linewidth]{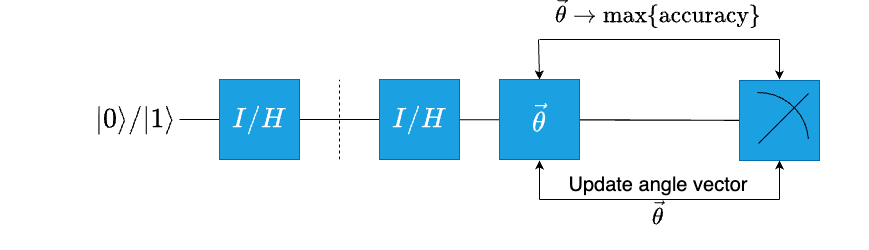}
\caption{}
\label{fig4a}
\end{subfigure}\hfill
\begin{subfigure}{\linewidth}
\centering
    \includegraphics[width=0.9\linewidth]{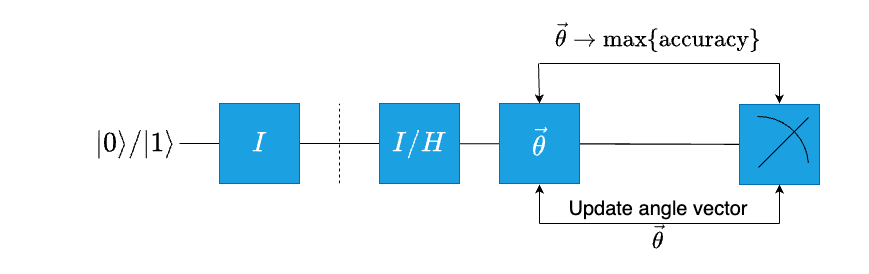}
\caption{}
\label{fig4b}
\end{subfigure}
\caption{Quantum circuits representing (a) QNN-BB84 and (b) QNN-B92 algorithms.}
\label{Fig4}
\end{figure*}
\begin{algorithm}[!t]
\DontPrintSemicolon
\nonl \textbf{Input:} Parameters $\theta$, Number of samples $S$, Number of key bits $N$ \\
\nonl \textbf{Output:} Metrics: Accuracy, Precision, Recall, F1-Score, QBER \\

\For{$s = 1$ to $S$}{
  Initialize empty lists: $K_b$, $K_m$, $F_k$ \\
  
  \For{$i = 1$ to $N$}{
    Sample a random bit $k \in \{0, 1\}$ \\
    \If{$k = 1$}{
      Prepare qubit in state $|1\rangle$ using $X$ gate
    }\Else{
      Prepare qubit in state $|0\rangle$
    }
    Append $k$ to $K_b$ \\
    
    Sample a random basis $l \in \{0, 1\}$ \\
    \If{$l = 1$}{
      Apply Hadamard gate to encode the basis
    }
    
    Sample a random measurement basis $m \in \{0, 1\}$ \\
    \If{$m = 1$}{
      Apply Hadamard gate to decode the basis
    }
    
    Apply a customized parameterized circuit with $\theta$ \\
    Measure the qubit and retrieve result $p$ \\
    Append $p$ to $K_m$ \\
    
    \If{$l = m$}{
      Append $p$ to $F_k$
    }
  }
  
  Compute metrics: Accuracy, Precision, Recall, F1-Score using $K_b$ and $K_m$ \\
  Compute QBER as $\text{QBER} = \frac{|F_k - K_b|}{|K_b|}$ \\
}
Compute averages and standard deviations for all metrics \\
Print ``QNN-BB84 Protocol" and metrics \\
\caption{Key Generation Process: QNN-BB84 Protocol}
\label{alg:qnn_bb84}
\end{algorithm}

\begin{algorithm}[!t]
\DontPrintSemicolon
\nonl \textbf{Input:} Parameters $\theta$, Number of samples $S$, Number of key bits $N$ \\
\nonl \textbf{Output:} Metrics: Accuracy, Precision, Recall, F1-Score, QBER \\

\For{$s = 1$ to $S$}{
  Initialize empty lists: $K_b$, $K_m$, $F_k$ \\
  
  \For{$i = 1$ to $N$}{
    Sample a random bit $k \in \{0, 1\}$ \\
    \If{$k = 1$}{
      Prepare qubit in state $|+\rangle$ using Hadamard gate
    }\Else{
      Prepare qubit in state $|0\rangle$
    }
    Append $k$ to $K_b$ \\
    
    Sample a random measurement basis $m \in \{0, 1\}$ \\
    \If{$m = 1$}{
      Apply Hadamard gate to decode the basis
    }
    
    Apply a customized parameterized circuit with $\theta$ \\
    Measure the qubit and retrieve result $p$ \\
    Append $p$ to $K_m$ \\
    
    \If{$k = m$}{
      Append $p$ to $F_k$
    }
  }
  
  Compute metrics: Accuracy, Precision, Recall, F1-Score using $K_b$ and $K_m$ \\
  Compute QBER as $\text{QBER} = \frac{|F_k - K_b|}{|K_b|} \cdot 100$ \\
}
Compute averages and standard deviations for all metrics \\
Print ``QNN-B92 Protocol" and metrics \\
\caption{Key Generation Process: QNN-B92 Protocol}
\label{alg:qnn_b92}
\end{algorithm}

\begin{figure*}[]
\centering
\begin{subfigure}{\linewidth}
\centering
    \includegraphics[width=0.9\linewidth]{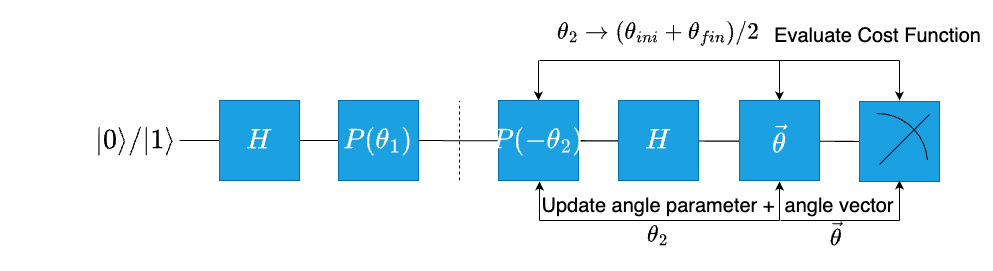}
\caption{}
\label{fig5a}
\end{subfigure}\hfill
\begin{subfigure}{\linewidth}
\centering
    \includegraphics[width=0.9\linewidth]{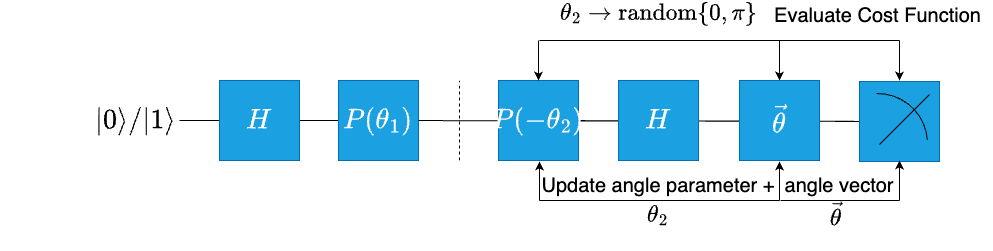}
\caption{}
\label{fig5b}
\end{subfigure}\hfill
\caption{Quantum circuits representing (a) QNN-QRL-V.1 and (b) QNN-QRL-V.2 algorithms.}
\label{Fig5}
\end{figure*}

\begin{algorithm}[!t]
\DontPrintSemicolon
\nonl \textbf{Input:} Parameters $\theta$, Number of samples $S$, Number of key bits $N$, Number of iterations $I$ \\
\nonl \textbf{Output:} Metrics: Accuracy, Precision, Recall, F1-Score, QBER \\

\For{$s = 1$ to $S$}{
  Initialize empty lists: $K_b$, $K_m$, $F_k$ \\

  \For{$i = 1$ to $N$}{
    Alice randomly selects a bit $k \in \{0, 1\}$ and encodes using $P(\theta_1)$, where $\theta_1 \in [0, \pi]$ \\
    Append $k$ to $K_b$ \\

    Bob initializes decoding bounds: $n_1 = 0$, $n_2 = \pi/2$, $n_3 = \pi$ \\
    Initialize Q-values: $Q(n_1) = Q(n_2) = Q(n_3) = 0$ \\

    \For{$j = 1$ to $I$}{
      \textbf{Environment:} Prepare qubit in $\ket{k}$ state, apply $H$, then $P(\theta_1)$ \\
      \textbf{Agent:} Bob selects decoding angle $\theta_2 \in \{n_1, n_2, n_3\}$ \\
      Apply $P(-\theta_2)$ and Hadamard gate, followed by parameterized circuit $U(\theta)$ \\
      Measure the qubit and compute probabilities $P_0$, $P_1$ \\
      \textbf{Reward:} $R = \max(P_0, P_1)$ \\
      \textbf{Q-Update:} $Q(\theta_2) \gets R$ \\
      Update interval $\{n_1, n_2, n_3\}$ toward the angle with highest $Q$ \\
      \If{converged (e.g., $|n_3 - n_1| < \epsilon$)}{break}
    }

    Set $\theta_2 = \arg\max Q(\theta)$ \\
    Prepare new qubit in $\ket{k}$, apply $H$, $P(\theta_1)$, $P(-\theta_2)$, and parameterized $U(\theta)$ \\
    Measure and append outcome to $K_m$ \\
  }

  Compare $K_b$ and $K_m$ to generate final key $F_k$ \\
  Compute metrics: Accuracy, Precision, Recall, F1-Score from $K_b$ and $K_m$ \\
  Compute QBER: $\text{QBER} = \frac{1}{N} \sum_{i=1}^N \mathbb{1}_{K_b[i] \ne K_m[i]} \cdot 100$ \\
}

Compute mean and standard deviation of all metrics across $S$ samples \\
Print ``QNN-QRL-V.1 Protocol" summary and final performance metrics \\
\caption{Key Generation Process: QNN-QRL-V.1 with QRL Agent Feedback}
\label{alg:qnn_qrl_v1}
\end{algorithm}

\begin{algorithm}[!t]
\DontPrintSemicolon
\nonl \textbf{Input:} Parameters $\theta$, Number of samples $S$, Number of key bits $N$, Number of iterations $I$ \\
\nonl \textbf{Output:} Metrics: Accuracy, Precision, Recall, F1-Score, QBER \\

\For{$s = 1$ to $S$}{
  Initialize empty lists: $K_b$, $K_m$, $F_k$ \\
  
  \For{$i = 1$ to $N$}{
    Alice selects a random bit $k \in \{0, 1\}$ \\
    Append $k$ to $K_b$ \\
    Select a random encoding angle $\theta_1 \in [0, \pi]$ \\

    Bob initializes search bounds: $n_1 = 0$, $n_2 = \texttt{rand}(0, \pi)$, $n_3 = \pi$ \\
    Initialize Q-values: $Q(n_1) = Q(n_2) = Q(n_3) = 0$ \\

    \For{$j = 1$ to $I$}{
      Prepare quantum state $\ket{k}$ using Hadamard and $P(\theta_1)$ gates \\
      
      Sample decoding angles: \\
      \Indp $m_1 = n_1 + \texttt{rand} \cdot (n_2 - n_1)$, \\
      $m_2 = n_2 + \texttt{rand} \cdot (n_3 - n_2)$ \\
      \Indm

      Apply $P(-m_1)$ and $P(-m_2)$ on copies of the quantum state \\
      Apply Hadamard gates followed by variational quantum circuit $U(\theta)$ \\
      Measure both circuits to obtain probabilities $P_0$ and $P_1$ \\
      
      \textbf{Reward:} $R = \max(P_0, P_1)$ \\
      \textbf{Agent Update:} $Q(m_1) \leftarrow R$ or $Q(m_2) \leftarrow R$ \\
      Update bounds $\{n_1, n_2, n_3\}$ to focus on high-reward region \\
      \If{convergence condition met}{break}
    }

    Select final decoding angle $\theta_2 = \arg\max Q$ \\
    Reconstruct the quantum state $\ket{k}$ and apply $H$, $P(\theta_1)$, $P(-\theta_2)$ \\
    Apply variational circuit $U(\theta)$ and perform measurement \\
    Append result to $K_m$ \\
  }

  Compare $K_b$ and $K_m$ to generate $F_k$ \\
  Compute Accuracy, Precision, Recall, F1-Score from $K_b$ and $K_m$ \\
  Compute QBER: $\text{QBER} = \frac{1}{N} \sum_{i=1}^{N} \mathbb{1}_{K_b[i] \ne K_m[i]} \cdot 100$ \\
}

Compute mean and standard deviation for all performance metrics \\
Print ``QNN-QRL-V.2 Protocol" summary and results \\
\caption{Quantum Key Generation Process: QNN-QRL-V.2 Protocol}
\label{alg:qrl_qkd_v2}
\end{algorithm}

\subsubsection{QRL-V.2.}
In this procedure, the above same process is repeated, with the only difference being the selection of $\theta$ values. In this scenario, the next $\theta$ is picked randomly rather than dividing the entire range by 2, so that $\Delta \theta$ converges around 1.2. The initial two ranges are also determined by randomly selecting an angle from the original range. The functions are calculated in both ranges, and the next range is carried forward. A random angle value chosen and the process is repeated until $\Delta \theta$ hits 1.2. This is the only difference from the QRL-V.1 algorithm. QRL-V.2 is another version of QRL-V.1, thus we refer to it as such. Algorithm \ref{alg:qkg_random_angles} provides a step-by-step process. Figure \ref{fig3b} depicts the quantum circuit for the entire algorithm. This technique uses the same quantum circuit as the previous QRL-V.1 algorithm; however, the learning strategy changes, involving the random selection of Bob's angle parameter, which ultimately determines the final bit received by Bob. The graph \ref{fig2c} shows that using this technique, after a number of iterations, $\Delta \theta$ reaches a value around 1.2, where $f1(\Delta \theta)$ bigger than $f2(\Delta \theta)$, which will help in choosing the bit transmitted by Alice.

\subsection{QNN Algorithms}
\subsubsection{QNN-BB84}
This is an integrated method of QNN and BB84 protocol. Mainly, in the BB84 protocol, the QNN algorithm has been integrated to improve its efficiency. The schematic diagram representing the algorithm is depicted in Fig. \ref{fig4a}. From the circuit, it is clear that Alice prepares $\ket{0}/\ket{1}$ states and randomly encodes them using Identity (I) or Hadamard (H) operations and sends them to Bob. After receiving the qubit, Bob randomly performs I or H operations. This completes the BB84 protocol, which follows for multiple number of sequences of bits. Now, the QNN algorithm is integrated, where a parametrized quantum circuit (PQC) is used, denoted by $\vec{\theta}$. This is the learning vector parameter in the quantum circuit that learns to maximize the accuracy of the overall algorithm, that is to collect the final key with the least QBER and the maximum matching with the initial key sent by Alice. The step-by-step process is given in Algorithm \ref{alg:qnn_bb84}. 

\subsubsection{QNN-B92}

This approach combines QNN and the B92 protocol. Specifically, the QNN algorithm has been implemented into the B92 protocol to improve its efficiency. Fig. \ref{fig4b} shows a schematic illustration of the algorithm. The quantum circuit displays Alice preparing $\ket{0}/\ket{1}$ states and randomly encoding them using the Identity (I) operation before sending them to Bob. After obtaining the qubit, Bob randomly performs I or H operations. This completes the B92 protocol, which is used for a multiple number of bit sequences. The QNN method is integrated using a PQC represented by $\vec{\theta}$. This is the learning vector parameter in the quantum circuit that learns to maximize the overall algorithm's accuracy, i.e., collect the final key with the lowest QBER and the closest match to Alice's generated key. Algorithm \ref{alg:qnn_b92} provides a step-by-step method.

\subsection{QNN-QRL Algorithms}
\subsubsection{QNN-QRL-V.1.}
This is an integrated method of QNN and QRL-V.1. protocol. Mainly, in the QRL-V.1 protocol, the QNN algorithm has been integrated to improve its efficiency. The schematic diagram representing the algorithm is depicted in Fig. \ref{fig5a}. In this case, the initial part of the quantum circuit is the same as that of the QRL-V.1. At the end of the circuit, the PQC is added to integrate the QNN algorithm with this method. The learning angle vector $\vec{\theta}$ learns the whole algorithm to find the optimal parameter such that the angle parameter $\theta_2$ gets the appropriate value, along with maximizing the overall accuracy of the whole algorithm. It is worth noticing that the algorithm learns the $\theta_2$ parameter of Bob (which comes from the QRL algorithm) and angle vector $\vec{\theta}$ (that comes from the QNN algorithm), hence integrating both QRL-V.1 and QNN algorithms into one protocol, leading to the QNN-QRL-V.1 algorithm. The strategy of choosing $\theta_2$ follows the QRL-V.1 algorithm, and the strategy for finding $\vec{\theta}$ follows optimizing the objective function that maximizes the accuracy of the overall algorithm. The step-by-step process is given in Algorithm \ref{alg:qnn_qrl_v1}.

\subsubsection{QNN-QRL-V.2.}
This is an integrated approach of QNN and QRL-V.2. protocol. Specifically, the QRL-V.2 protocol incorporates the QNN algorithm to boost its efficiency. Fig. \ref{fig5b} shows a schematic illustration of the algorithm. In this scenario, the quantum circuit's initial component is identical to that of the QRL-V.2. The difference is that at the end of the circuit, a PQC is introduced to integrate the QNN algorithm into this technique. The learning angle vector $\vec{\theta}$ learns the whole algorithm to find the optimal parameter so that the angle parameter $\theta_2$ gets the appropriate value, while also maximizing the overall accuracy of the algorithm. The algorithm learns Bob's $\theta_2$ parameter (from the QRL algorithm) and angle vector $\vec{\theta}$ (from the QNN algorithm), combining the QRL-V.2 and QNN algorithms into a single protocol, resulting in the QNN-QRL-V.2 algorithm. The approach for selecting $\theta_2$ is based on the QRL-V.2 algorithm, which randomly selects a value from a specific interval with each iteration. The strategy for determining $\vec{\theta}$ is based on maximizing the objective function to optimize the algorithm's accuracy. Algorithm \ref{alg:qrl_qkd_v2} provides a step-by-step process.

\begin{table*}[]
    \centering
    \begin{tabular}{|c|c|c|c|c|c|}
    \hline
    \textbf{Protocols} &  \textbf{Accuracy} & \textbf{Precision} & \textbf{Recall} & \textbf{F1 Score} & \textbf{QBER}\\
    \hline
    BB84 & 0.755 $\pm$ 0.026 &  0.741 $\pm$ 0.046 & 0.755 $\pm$ 0.061 & 0.747 $\pm$ 0.045 & 0.498 $\pm$ 0.035\\
    \hline
    B92 & 0.528 $\pm$ 0.046 &  0.496 $\pm$ 0.096 & 0.262 $\pm$ 0.063 & 0.341 $\pm$ 0.071 & 0.483 $\pm$ 0.050\\
    \hline
    QRL-QKD-V.1 & 0.478 $\pm$ 0.048 &  0.488 $\pm$ 0.068 & 0.467 $\pm$ 0.068 & 0.476 $\pm$ 0.065 & 0.522 $\pm$ 0.048\\
    \hline
    QRL-QKD-V.2 & 0.476 $\pm$ 0.042 &  0.490 $\pm$ 0.087 & 0.435 $\pm$ 0.067 & 0.458 $\pm$ 0.068 & 0.524 $\pm$ 0.042\\
    \hline
    QNN-BB84 & 1.000 $\pm$ 0.000 &  1.000 $\pm$ 0.000 & 1.000 $\pm$ 0.000 & 1.000 $\pm$ 0.000 & 0.493 $\pm$ 0.048\\
    \hline
    QNN-B92 & 0.472 $\pm$ 0.051 & 0.469 $\pm$ 0.051 & 0.996 $\pm$ 0.008 & 0.636 $\pm$ 0.046 & 0.500 $\pm$ 0.038\\
    \hline
    QNN-QRL-V.1 & 0.984 $\pm$ 0.009 & 0.973 $\pm$ 0.022 & 0.991 $\pm$ 0.011 & 0.982 $\pm$ 0.012 & 0.016 $\pm$ 0.009\\
    \hline
    QNN-QRL-V.2 & 0.987 $\pm$ 0.012 & 0.990 $\pm$ 0.014 & 0.985 $\pm$ 0.015 & 0.987 $\pm$ 0.013 & 0.013 $\pm$ 0.002\\
    \hline
    \end{tabular}
    \caption{Evaluation metrics for the existing and proposed algorithms}
    \label{tab:evaluation_metrics}
\end{table*}
\begin{figure*}[]
\begin{subfigure}{0.5\linewidth}
    \includegraphics[width=\linewidth]{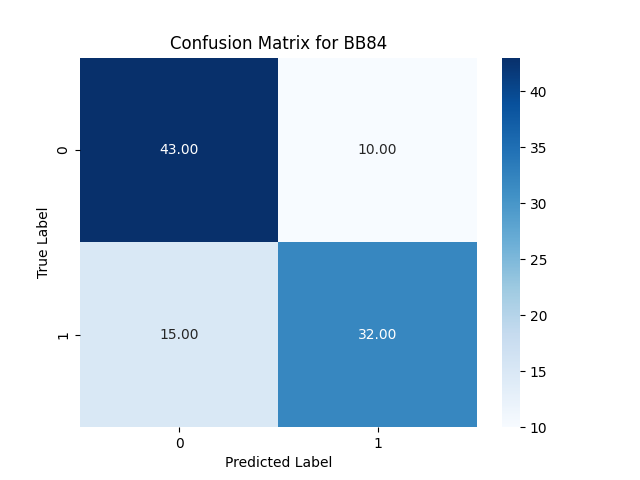}
    \caption{}
    \label{cnn2x2lisa}
\end{subfigure}\hfill
\begin{subfigure}{0.5\linewidth}
     \includegraphics[width=\linewidth]{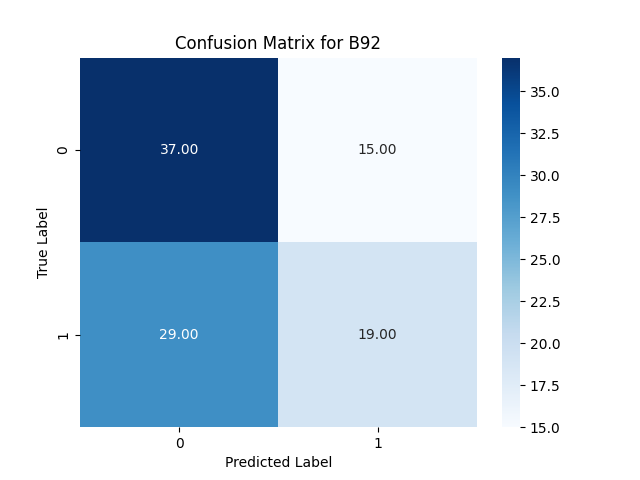}
    \caption{}
    \label{cnn4x4lisa}
\end{subfigure}\hfill
\begin{subfigure}{0.5\linewidth}
     \includegraphics[width=\linewidth]{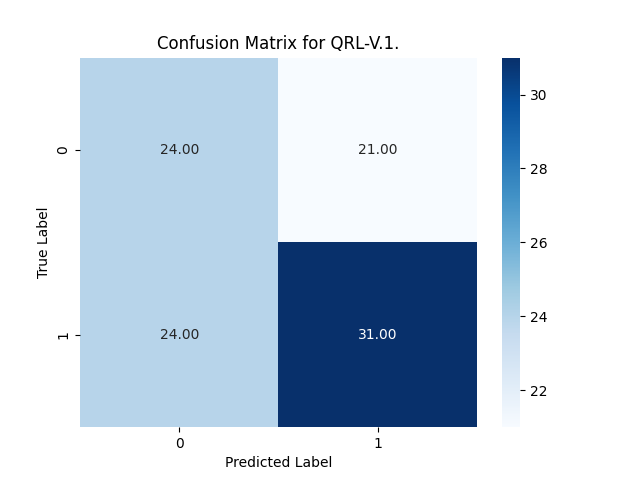}
    \caption{}
    \label{cnn4x4lisa}
\end{subfigure}\hfill
\begin{subfigure}{0.5\linewidth}
     \includegraphics[width=\linewidth]{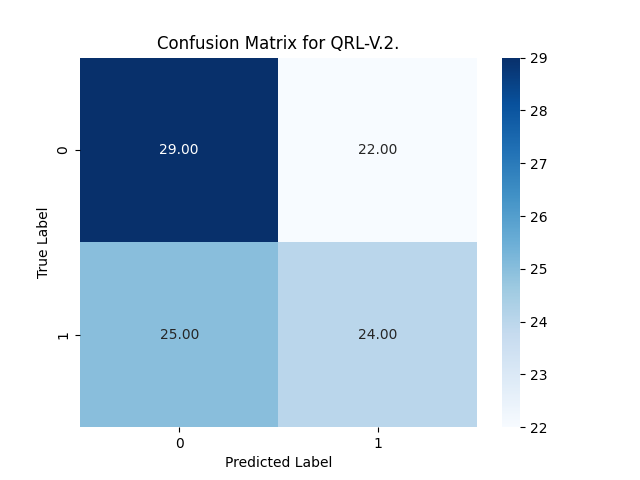}
    \caption{}
    \label{cnn4x4lisa}
\end{subfigure}
\caption{Confusion matrices for the algorithms BB84, B92, QRL-V.1., and QRL-V.2.}
\label{confusion_matrices}
\end{figure*}

\begin{figure*}[]
\begin{subfigure}{0.5\linewidth}
    \includegraphics[width=\linewidth]{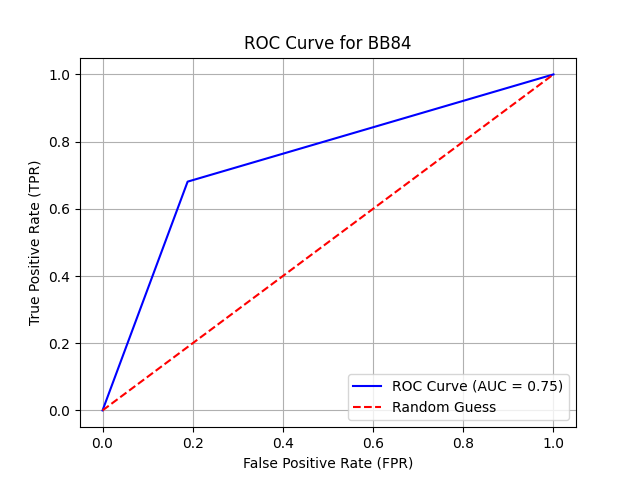}
    \caption{}
    \label{cnn2x2lisa}
\end{subfigure}\hfill
\begin{subfigure}{0.5\linewidth}
     \includegraphics[width=\linewidth]{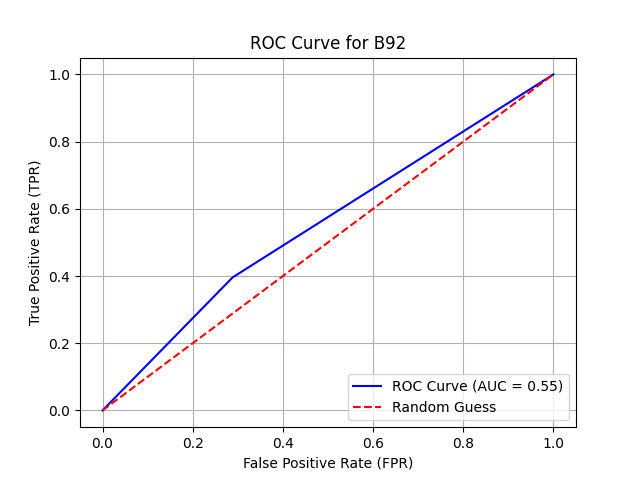}
    \caption{}
    \label{cnn4x4lisa}
\end{subfigure}\hfill
\begin{subfigure}{0.5\linewidth}
     \includegraphics[width=\linewidth]{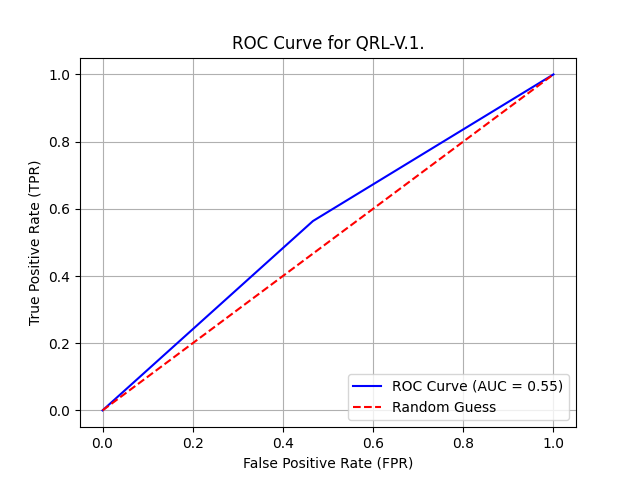}
    \caption{}
    \label{cnn4x4lisa}
\end{subfigure}\hfill
\begin{subfigure}{0.5\linewidth}
     \includegraphics[width=\linewidth]{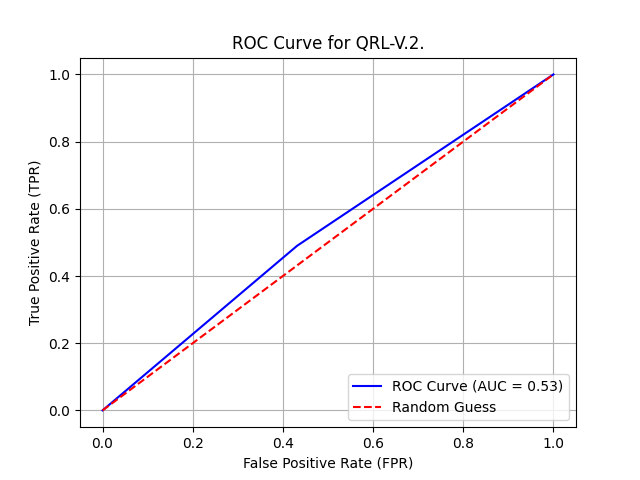}
    \caption{}
    \label{cnn4x4lisa}
\end{subfigure}
\caption{ROC curves for the algorithms BB84, B92, QRL-V.1., and QRL-V.2.}
\label{roc_curves}
\end{figure*}
\begin{figure*}
\centering
\begin{subfigure}{0.5\linewidth}
\centering
    \includegraphics[width=\linewidth]{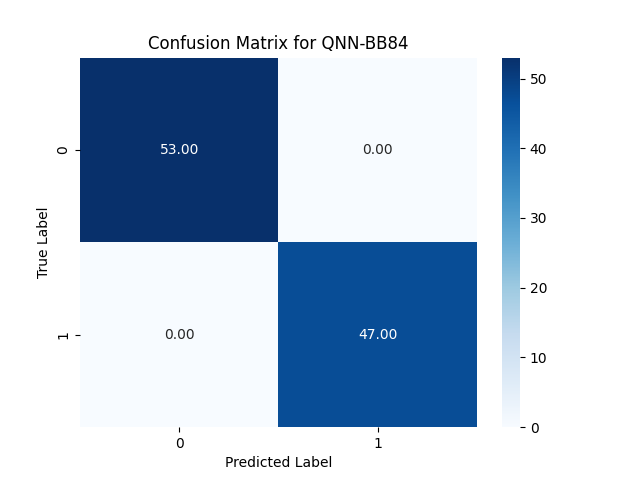}
    \caption{}
    \label{cnn2x2lisa}
\end{subfigure}\hfill
\begin{subfigure}{0.5\linewidth}
\centering
     \includegraphics[width=\linewidth]{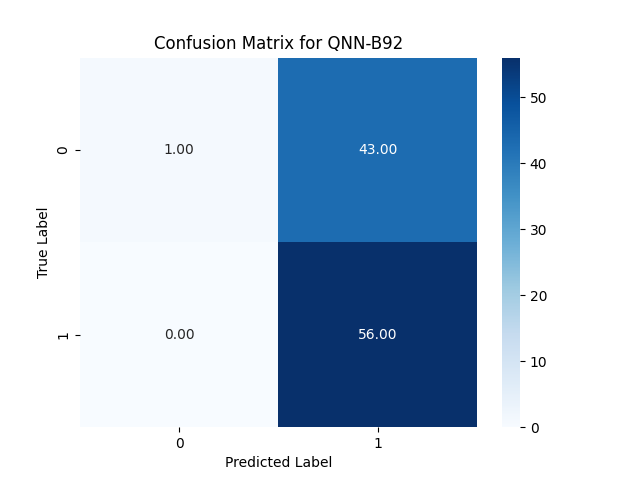}
    \caption{}
    \label{cnn4x4lisa}
\end{subfigure}\hfill
\begin{subfigure}{0.5\linewidth}
\centering
     \includegraphics[width=\linewidth]{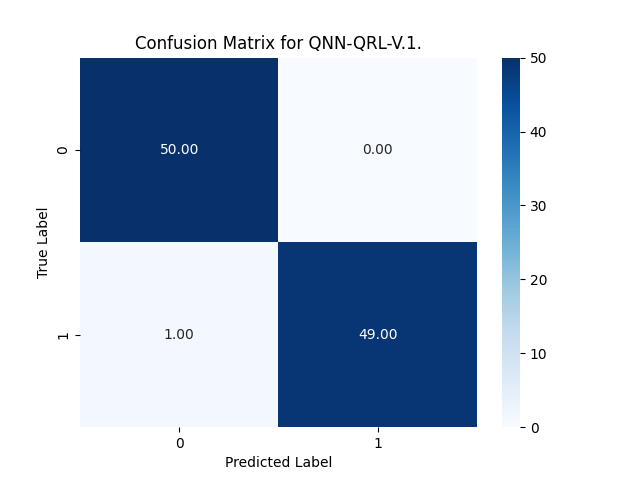}
    \caption{}
    \label{cnn4x4lisa}
\end{subfigure}\hfill
\begin{subfigure}{0.5\linewidth}
\centering
     \includegraphics[width=\linewidth]{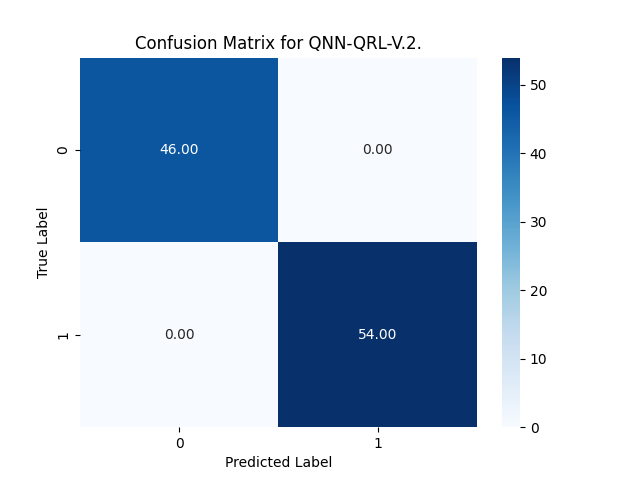}
    \caption{}
    \label{cnn4x4lisa}
\end{subfigure}\hfill
\caption{Confusion matrices for the QNN integrated algorithms BB84, B92, QRL-V.1., and QRL-V.2.}
\label{confusionmatricesfig8}
\end{figure*}

\begin{figure*}[]
\begin{subfigure}{0.5\linewidth}
    \includegraphics[width=\linewidth]{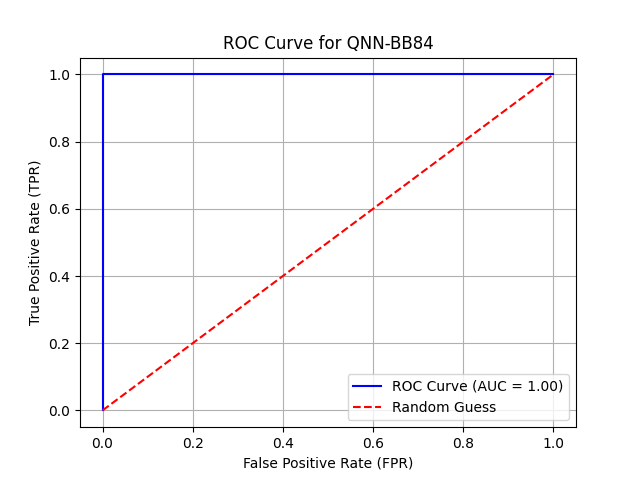}
    \caption{}
    \label{cnn2x2lisa}
\end{subfigure}\hfill
\begin{subfigure}{0.5\linewidth}
     \includegraphics[width=\linewidth]{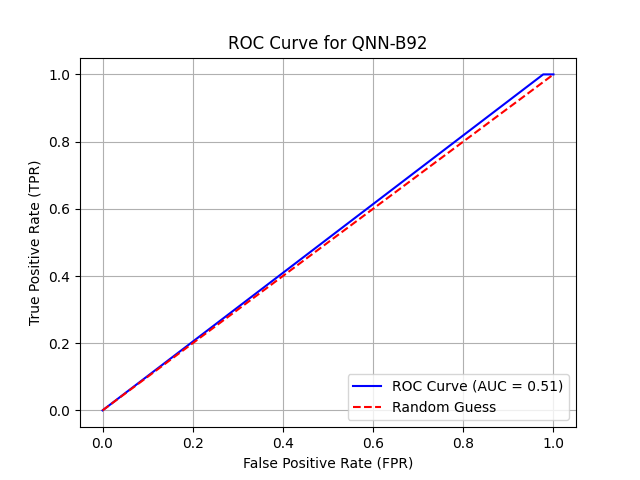}
    \caption{}
    \label{cnn4x4lisa}
\end{subfigure}\hfill
\begin{subfigure}{0.5\linewidth}
     \includegraphics[width=\linewidth]{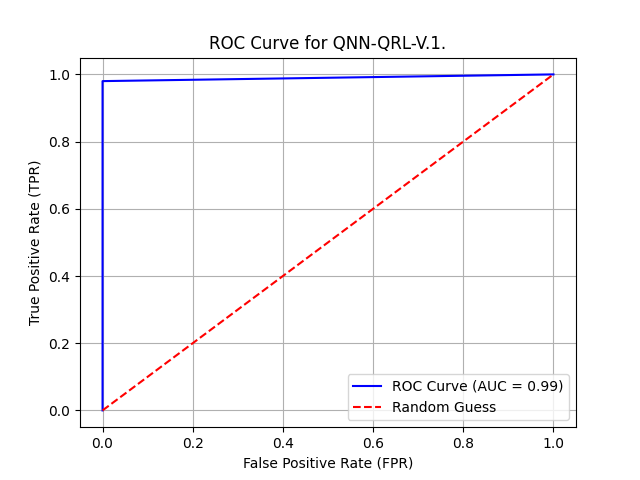}
    \caption{}
    \label{cnn4x4lisa}
\end{subfigure}\hfill
\begin{subfigure}{0.5\linewidth}
     \includegraphics[width=\linewidth]{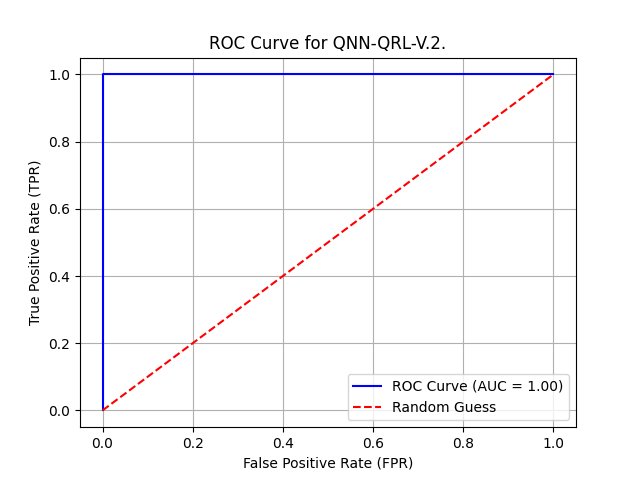}
    \caption{}
    \label{cnn4x4lisa}
\end{subfigure}
\caption{ROC curves for the QNN integrated algorithms BB84, B92, QRL-V.1., and QRL-V.2.}
\label{roc_curves_fig7}
\end{figure*}
\begin{figure*}[]
\begin{subfigure}{0.5\linewidth}
    \includegraphics[width=\linewidth]{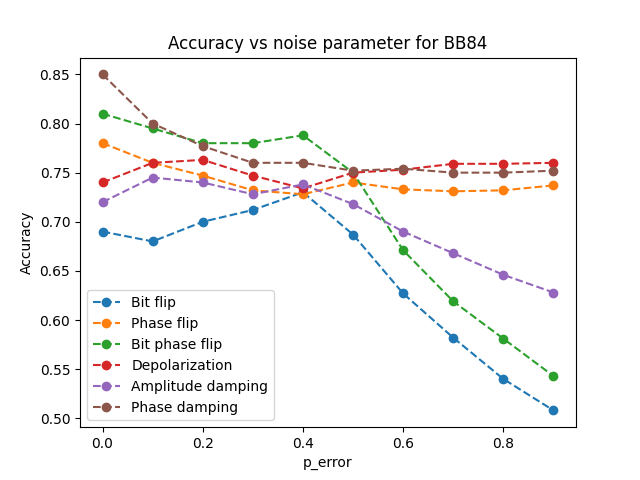}
    \caption{}
    \label{fig:noise_bb84}
\end{subfigure}\hfill
\begin{subfigure}{0.5\linewidth}
     \includegraphics[width=\linewidth]{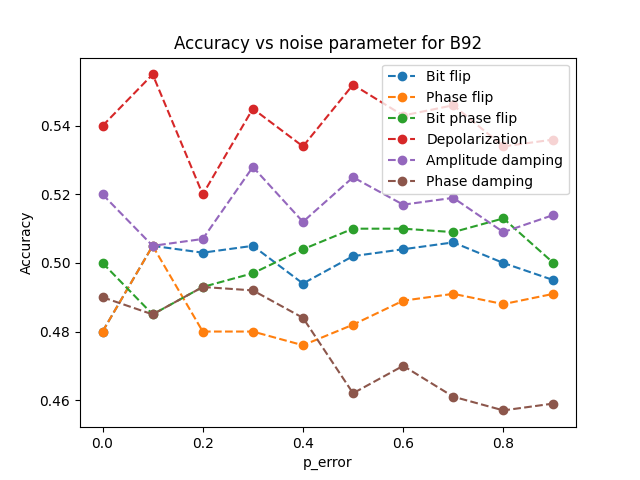}
    \caption{}
    \label{fig:noise_b92}
\end{subfigure}\hfill
\begin{subfigure}{0.5\linewidth}
     \includegraphics[width=\linewidth]{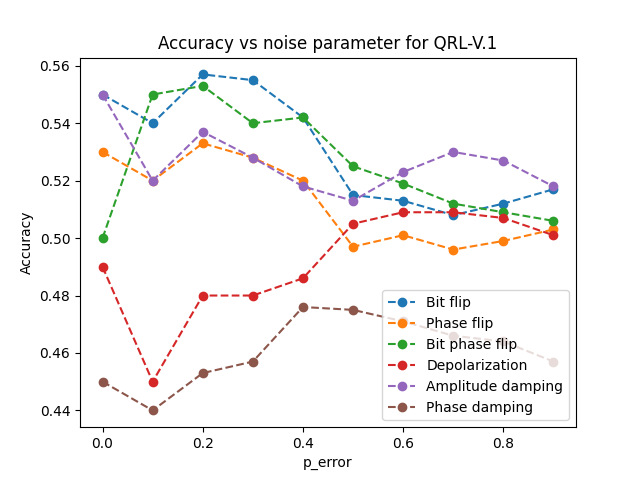}
    \caption{}
    \label{fig:noise_qrl-v1}
\end{subfigure}\hfill
\begin{subfigure}{0.5\linewidth}
     \includegraphics[width=\linewidth]{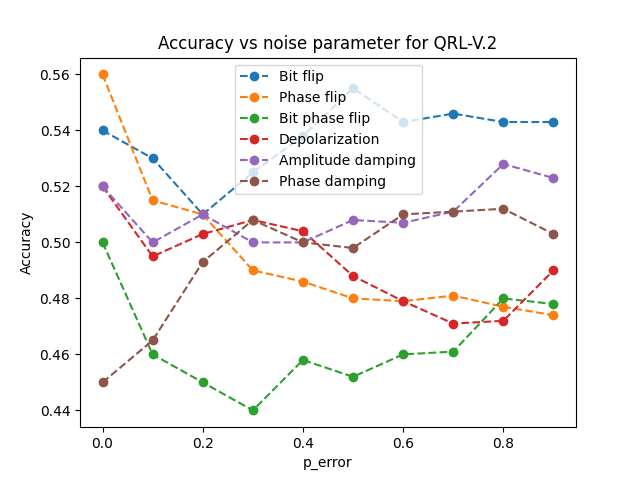}
    \caption{}
    \label{fig:noise_qrl-v2}
\end{subfigure}
\caption{Noise plots for a) BB84, b) B92, c) QRL-V.1 and d) QRL-V.2.}
\label{noiseplots_bb84_qrlv2}
\end{figure*}

\begin{figure*}[]
\begin{subfigure}{0.5\linewidth}
    \includegraphics[width=\linewidth]{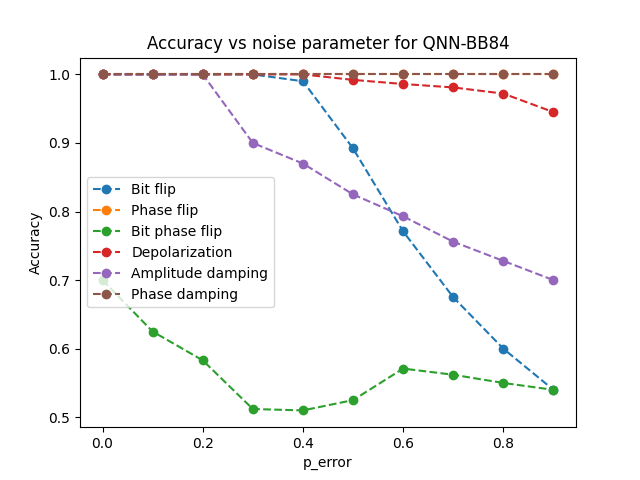}
    \caption{}
    \label{fig:noise_qnn-bb84}
\end{subfigure}\hfill
\begin{subfigure}{0.5\linewidth}
     \includegraphics[width=\linewidth]{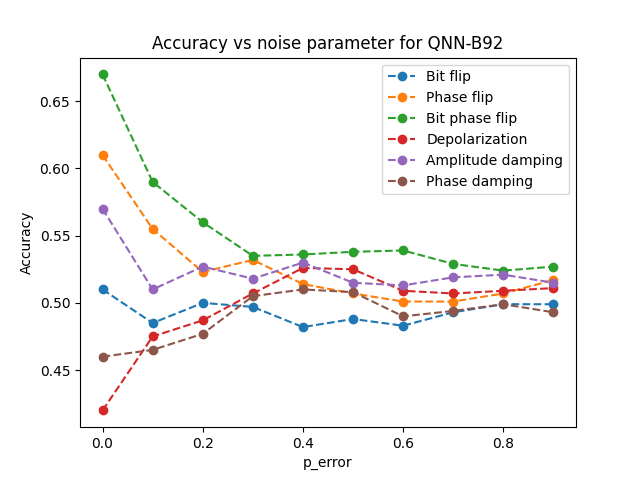}
    \caption{}
    \label{fig:noise_qnn-b92}
\end{subfigure}\hfill
\begin{subfigure}{0.5\linewidth}
     \includegraphics[width=\linewidth]{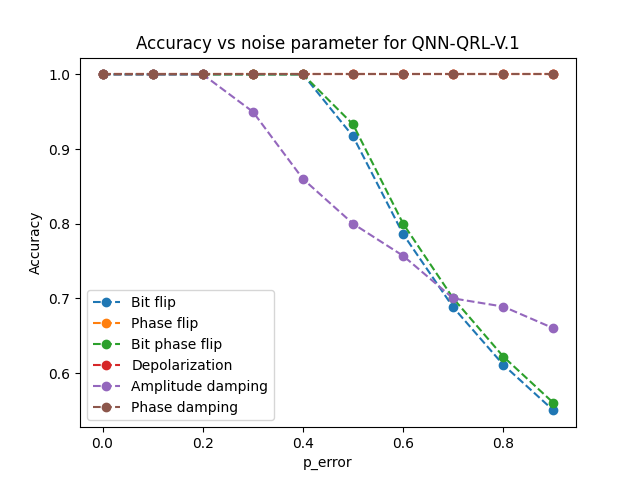}
    \caption{}
    \label{fig:noise_qnn-qrl-v1}
\end{subfigure}\hfill
\begin{subfigure}{0.5\linewidth}
     \includegraphics[width=\linewidth]{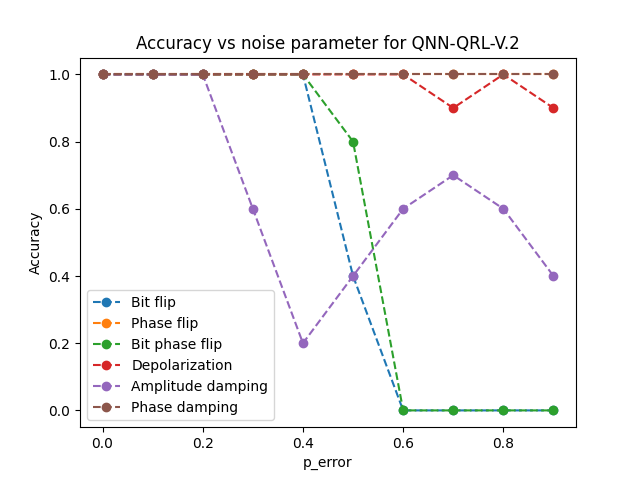}
    \caption{}
    \label{fig:noise_qnn-qrl-v2}
\end{subfigure}
\caption{Noise plots for a) QNN-BB84, b) QNN-B92, c) QNN-QRL-V.1 and d) QNN-QRL-V.2.}
\label{noiseplots_qnn-bb84_qrlv2}
\end{figure*}

\section{Results\label{results}}

\iffalse
\subsection{Settings}
The study has the following objectives:

\begin{enumerate}
\item To develop QRL-based QKD protocols and compare their performances with existing protocols.
\item To integrate the QNN algorithm into the existing and proposed QRL-based QKD protocols and compare their performances. 
\item To compare the quality of the key generation process, and minimize the QBER for the proposed QNN-based algorithms.
\end{enumerate}
To achieve 1), QRL-based QKD protocols such as QRL-V.1 and QRL-V.2 are developed and their evaluation metrics are compared with the existing protocols. Then for 2), the QNN algorithm is integrated with both the existing and proposed QKD protocols to propose novel QKD algorithms to significantly improve their performance over the other algorithms. For checking the quality of the key generation process, evaluation metrics such as accuracy, precision, recall, F1 score and QBER are calculated and optimized for each QNN-based algorithm.
\fi
\subsection{Datasets, Preprocessing and Hyperparameters}
The dataset mainly involves the key generation process which is performed using random generation of bits by Alice and sent to Bob for further processing. The algorithm supports the generation keys of length 100 to 1000-bit strings for collecting faster results. There are no preprocessing steps involved however, the key generation process uses a random value between 0 and 1, and accordingly decides the quantum operations for encoding and decoding in the quantum circuit involved in the respective protocols. The QISKit platform is used for the experiment, where an `automatic' simulator is used with 1024 numbers of shots. For the optimization of the circuit parameters, a Cobyla optimizer is used. The number of iterations taken for QNN integrated algorithms is 5, however, in the first iteration itself, the best results are achieved.

\subsection{Noise Models}
Investigating various noise models, including bit flip, phase flip, bit-phase flip, depolarizing, amplitude damping, and phase damping, is crucial for understanding the robustness and dependability of quantum algorithms in real-world circumstances \cite{bikash_noise}. These noise models depict different types of quantum errors caused by interactions with the environment. For example, bit flip errors change the state of a qubit, whereas phase flip errors affect the phase of the quantum state. Depolarizing noise causes random mistakes, whereas amplitude damping simulates energy loss, a major problem with superconducting qubits. Phase damping describes the loss of quantum coherence without requiring energy dissipation. 

%%%%%%%%%%%%%%%%%%%%%%%%%%%%%%%%%
\subsection{Results\label{sec5}}
%%%%%%%%%%%%%%%%%%%%%%%%%%%%%%%%%%%%%%%%%%%%
From Table \ref{tab:evaluation_metrics}, it is observed that the BB84 protocol achieves a notably higher performance compared to the B92 protocol. Specifically, BB84 attains an accuracy of $0.755 \pm 0.026$, whereas B92 records a lower accuracy of $0.528 \pm 0.046$. Moreover, BB84 outperforms B92 in all other evaluation metrics including precision, recall, and F1 score. However, the QBER value for BB84 is $0.498 \pm 0.035$, which is marginally higher than that of B92, which stands at $0.483 \pm 0.050$. The QRL-based algorithms, namely QRL-QKD-V.1 and QRL-QKD-V.2, yield lower accuracy scores of $0.478 \pm 0.048$ and $0.476 \pm 0.042$ respectively, slightly below that of the B92 protocol. Nonetheless, their precision, recall, and F1 scores exhibit moderate improvements compared to B92, albeit with higher QBER values of $0.522 \pm 0.048$ and $0.524 \pm 0.042$. Among all methods, the QNN-integrated algorithms demonstrate substantial performance gains. QNN-BB84 achieves perfect scores of $1.000 \pm 0.000$ across all evaluation metrics, accuracy, precision, recall, and F1 score while maintaining a QBER of $0.493 \pm 0.048$, which is slightly lower than that of the classical BB84 protocol. The QNN-QRL variants, QNN-QRL-V.1 and QNN-QRL-V.2, also show near-perfect performance with accuracies of $0.984 \pm 0.009$ and $0.987 \pm 0.012$ respectively, and QBER values as low as $0.016 \pm 0.009$ and $0.013 \pm 0.002$, the lowest among all protocols tested. Although QNN-B92 does not achieve competitive accuracy ($0.472 \pm 0.051$), it demonstrates an exceptionally high recall of $0.996 \pm 0.008$, leading to a relatively improved F1 score of $0.636 \pm 0.046$. However, its QBER remains comparable to other protocols at $0.500 \pm 0.038$. These results collectively affirm that the integration of QNN significantly enhances the performance of QKD protocols, particularly in terms of classification accuracy and error rates.

\iffalse
The confusion matrices for the algorithms BB84, B92, QRL-V.1, and QRL-V.2 are given in Fig. \ref{confusion_matrices}. It can be observed that the amount of true classification is 70\%, 60\%, 40\%, and 70\% for the BB84, B92, QRL-V.1, and QRL-V.2 algorithms respectively. BB84 and QRL-V.2. are the best performers among them, while QRL-V.2 is the worst performer among them. This suggests the amount of misclassification of keys is the highest in the case of QRL-V.1 as compared to the other algorithms, leading to 60\%. It can be noted that the amount of misclassification of one of the classes for B92 is completely zero. In Fig. \ref{roc_curves}, the ROC curves of the above algorithms are shown. The AUC scores of BB84, B92, QRL-V.1, and QRL-V.2 are 0.71, 0.67, 0.62, and 0.75. It can be noted that the AUC score of QRL-V.2 is the best among the above algorithms, which is followed by BB84, and B92, while QRL-V.1 has the lowest AUC score. This means the true positive rate in the case of QRL-V.2 is the highest leading to one of the most feasible and best methods among these algorithms. In other words, the QRL-V.2 has outperformed the other algorithm in terms of AUC score.
\fi

The confusion matrices for the algorithms BB84, B92, QRL-V.1, and QRL-V.2 are shown in Fig.~\ref{confusion_matrices}. From the matrices, we observe that BB84 and QRL-V.2 achieve the highest number of true classifications (BB84: 43 + 32, QRL-V.2: 29 + 24), indicating better performance than B92 and QRL-V.1. Specifically, B92 shows a total of 56 correct classifications (37 + 19), while QRL-V.1 records the lowest at 52 (24 + 28). This suggests that QRL-V.1 suffers from the highest misclassification rate among the four. Interestingly, B92 shows a completely correct classification of the majority class (label 0), with zero misclassified samples of that class. The ROC curves for the same set of algorithms are provided in Fig.~\ref{roc_curves}. The AUC scores for BB84, B92, QRL-V.1, and QRL-V.2 are approximately 0.75, 0.55, 0.55, and 0.53, respectively. BB84 demonstrates the best AUC performance, suggesting a higher true positive rate across threshold settings, while the QRL variants lag behind in discriminative power. In particular, QRL-V.2 and QRL-V.1 show near-random classification behavior with AUC scores close to 0.5. Thus, in terms of both classification accuracy and AUC, BB84 emerges as the most robust algorithm among the four evaluated.

\iffalse
Fig. \ref{confusionmatricesfig8} illustrates the confusion matrices for all QNN integrated algorithms, including QNN-BB84, QNN-B92, QNN-QRL-V.1, and QNN-QRL-V.2. The QNN-BB84, QNN-B92, QNN-QRL-V.1, and QNN-QRL-V.2 algorithms achieve 100\%, 60\%, 100\%, and 100\% true classification, respectively. QNN-BB84, QNN-QRL-V.1, and QRL-V.2 outperform them all, with QNN-B92 placing the worst. QNN-B92 has the largest rate of key misclassification compared to other algorithms, resulting in 40\%. It should be emphasized that the other algorithms generate no misclassification. Fig. \ref{roc_curves_fig7} depicts the ROC curves for the above algorithms. The AUC values for QNN-BB84, QNN-B92, QNN-QRL-V.1, and QNN-QRL-V.2 are respectively 1.0, 0.5, 1.0, and 1.0. It is worth noting that QNN-BB84, QNN-QRL-V.1, and QNN-QRL-V.2 have the highest AUC scores among the algorithms mentioned above, while QNN-B92 has the lowest. This suggests that QNN-BB84, QNN-QRL-V.1, and QNN-QRL-V.2 have the highest true positive rates, making them one of the most viable and effective approaches among all proposed and current algorithms.
\fi
%%%%%%%%%%%%%%%%%%%%%%%%%%%%%%%%%%%%%%%%%%%%
Fig.~\ref{confusionmatricesfig8} presents the confusion matrices for the QNN-integrated algorithms: QNN-BB84, QNN-B92, QNN-QRL-V.1, and QNN-QRL-V.2. Among these, QNN-BB84, QNN-QRL-V.1, and QNN-QRL-V.2 demonstrate almost perfect classification performance, with only one misclassified sample each in QNN-QRL-V.1 and QNN-QRL-V.2, and no misclassifications in QNN-BB84. In contrast, QNN-B92 shows a relatively lower performance, with 14 misclassifications (1 false positive and 13 false negatives), highlighting its weaker discriminative ability. This confirms that QNN-BB84, QNN-QRL-V.1, and QNN-QRL-V.2 are highly effective in accurate key classification, while QNN-B92 struggles significantly in comparison. Fig.~\ref{roc_curves_fig7} shows the ROC curves corresponding to these algorithms. The AUC scores are 1.00 for QNN-BB84, 0.51 for QNN-B92, 0.99 for QNN-QRL-V.1, and 1.00 for QNN-QRL-V.2. These results clearly indicate that QNN-BB84, QNN-QRL-V.1, and QNN-QRL-V.2 are superior in terms of true positive rate and overall classification quality. QNN-B92, with an AUC close to 0.5, performs no better than random guessing. Hence, the QNN-based integration significantly enhances the classification capability of all algorithms except QNN-B92, making QNN-BB84, QNN-QRL-V.1, and QNN-QRL-V.2 the most viable and effective models for secure quantum key distribution systems.

\subsection{Noisy Results}
From Fig. \ref{fig:noise_bb84}, it can be observed that the accuracy performance is greatly impacted by the bit-flip noise model, where the accuracy has dropped from 0.7 to 0.5. It is interesting to note that, around the noise parameter of near 0.4 value, the accuracy increases to a height of near 0.75. Then, next, it is impacted by the bitphaseflip noise model, where the accuracy drops from 0.8 to 0.55 showing a similar peak around near 0.4 noise parameter value. Next, amplitude damping noise affects the most, whereas the rest of the models fluctuate in a similar range throughout noise parameters. Hence, the BB84 protocol shows robustness against the other rest noise models. From Fig. \ref{fig:noise_b92}, it can be observed that there is no significant decrease in the accuracy of the B92 protocol for all noise models. So in a sense, it can be said that the B92 protocol is the most effective against all noise models. It can be mentioned that some of the noise models show an increase in accuracy with an increase in noise parameters, specifically bit flip, phase flip, and bit phase flip, showing a consistent increase in accuracy with an increase in noise parameters, while some others show some increment in accuracy then dropping the accuracy, such as depolarizing, amplitude damping, and phase damping. From Fig. \ref{fig:noise_qrl-v1}, it can be observed that for noise models such as bit flip, phase flip, bit phase flip, and amplitude damping, the accuracy fluctuates between 0.5 and 0.56. On the other hand, for depolarizing and phase damping, the accuracy fluctuates between 0.44 and 0.5. It is worth noting that both the depolarizing and phase-damping accuracies are increasing with the increase of noise parameters. From Fig. \ref{fig:noise_qrl-v2}, it can be observed that all the noise models vary between 0.44 and 0.56. In the phase flip and phase damping noise models, the accuracies are steadily decreasing and increasing, respectively.

In Fig. \ref{noiseplots_qnn-bb84_qrlv2}, the accuracies for QNN integrated models against different noisy channels are presented. 
From Fig. \ref{fig:noise_qnn-bb84}, it is observed that the accuracy of QNN-BB84 against phase flip, depolarizing, and phase damping noise remains almost constant and not greatly impacted by the noise parameters. Whereas, the accuracies for bit flip, bit phase flip, and amplitude damping are greatly impacted by the noise parameters reaching around 0.55 for bit flip and bit phase flip, whereas the accuracy for amplitude damping is reaching near 0.7. From Fig. \ref{fig:noise_qnn-b92}, it is clear that all the noise models' accuracies are saturating around 0.5 irrespective of their starting points. It is worth noting that phase damping and depolarizing accuracies are increasing from the starting point, whereas the rest are decreasing till they saturate. From Fig. \ref{fig:noise_qnn-qrl-v1}, it can be seen that the phase flip, depolarizing, and phase damping accuracies stay constant throughout the noise parameter range, while bit flip, bit phase flip, and amplitude damping are declining after a certain noise parameter. For example, bit flip and bit phase flip accuracies decline after 0.4 noise parameters and reach 0.5 accuracy, whereas the accuracy for amplitude damping declines after 0.2 noise parameters and reaches around 0.7. In Fig. \ref{fig:noise_qnn-qrl-v2}, the phase flip, bit phase flip, and depolarizing behave similarly to the previous one, while the bit flip, bit phase flip, and amplitude damping have a little different behaviour. The bit flip and bit phase flip accuracies decline after 0.4 and reach 0 accuracy after 0.6 noise parameter. The amplitude damping accuracy declines after 0.2 and then after 0.4, again increasing the accuracy to 0.4 at the maximum noise.

%%%%%%%%%%%%%%%%%%%%%%%%%%%%%%%%%%%%%%%%%%%%

\subsection{Discussion}

The reported results highlight the comparative performance of classical QKD protocols, QRL-based models, and QNN-integrated algorithms in the quantum key generation context. From Table~\ref{tab:evaluation_metrics}, it is evident that the BB84 protocol significantly outperforms B92 across all evaluation metrics. BB84 achieves an accuracy of $0.755 \pm 0.026$, surpassing B92's $0.528 \pm 0.046$, and similarly shows superior precision, recall, and F1-score. However, BB84’s QBER is $0.498 \pm 0.035$, which is slightly higher than B92's $0.483 \pm 0.050$, indicating a minor trade-off in terms of bit error rates. The QRL-based protocols, namely QRL-QKD-V.1 and QRL-QKD-V.2, report slightly lower accuracies than B92, at $0.478 \pm 0.048$ and $0.476 \pm 0.042$ respectively. Despite this, they achieve better performance in precision, recall, and F1-score than B92, suggesting that the QRL approach enhances classification quality. Nevertheless, their higher QBER values, $0.522 \pm 0.048$ for QRL-QKD-V.1 and $0.524 \pm 0.042$ for QRL-QKD-V.2 imply that these algorithms may introduce greater noise in key generation and require further optimization. The confusion matrices in Fig.~\ref{confusion_matrices} reinforce these observations. BB84 and QRL-V.2 attain the highest number of correct classifications (75 each), while B92 and QRL-V.1 perform less effectively, with QRL-V.1 yielding the lowest count at 52. Notably, B92 achieves perfect classification for the majority class (label 0), although its performance on the minority class is weaker. Corresponding ROC curves in Fig.~\ref{roc_curves} show that BB84 has the highest AUC (approximately 0.75), while QRL-V.1 and QRL-V.2 have AUCs of about 0.55 and 0.53, respectively near random-guessing thresholds underscoring BB84’s reliability.

Substantial performance improvements are observed with QNN integration. QNN-BB84 achieves perfect accuracy, precision, recall, and F1-score ($1.000 \pm 0.000$), with a slightly reduced QBER of $0.493 \pm 0.048$ compared to its classical counterpart. QNN-QRL-V.1 and QNN-QRL-V.2 also show near-perfect performance, with accuracies of $0.984 \pm 0.009$ and $0.987 \pm 0.012$ respectively, and exceptionally low QBERs of $0.016 \pm 0.009$ and $0.013 \pm 0.002$ the lowest among all methods. As seen in Fig.~\ref{confusionmatricesfig8}, QNN-BB84 yields flawless classification with no misclassifications. QNN-QRL-V.1 and QNN-QRL-V.2 misclassify only a single sample each, highlighting their robustness. In contrast, QNN-B92 misclassifies 14 instances (including both false positives and false negatives), which aligns with its lower accuracy of $0.472 \pm 0.051$. Although QNN-B92 achieves an outstanding recall of $0.996 \pm 0.008$, its overall performance remains subpar. Fig.~\ref{roc_curves_fig7} displays the ROC curves for the QNN-integrated models. QNN-BB84 and QNN-QRL-V.2 exhibit perfect AUC scores of 1.00, while QNN-QRL-V.1 achieves an AUC of 0.99, indicating excellent discriminative ability. Conversely, QNN-B92’s AUC of 0.51 is close to random guessing, underscoring its limitations. Overall, the results clearly demonstrate that QNN integration dramatically enhances QKD protocol performance both in terms of classification accuracy and error minimization. QNN-BB84, QNN-QRL-V.1, and QNN-QRL-V.2 are the most promising models, achieving near-ideal performance and positioning themselves as strong candidates for practical and secure quantum key distribution. However, the performance drop in QNN-B92 suggests that QNN enhancements may not generalize uniformly across all classical QKD schemes, warranting further protocol-specific investigation.

\section{Conclusion \label{conclusion}}
In this paper, we address practical limitations in QKD by using QML methods. We proposed two QRL-based algorithms, and improved the existing BB84 and B92 protocols with QNN integration, yielding four new algorithm variants. Comparative evaluation show that the proposed algorithms outperform existing algorithms in terms of accuracy, precision, recall, and F1 score. Integrating QRL and QNN into QKD protocols not only increases key generation rates but also reduces the QBER. The robustness of the above models is investigated against six different noisy channels. By comparing our algorithms to these models, we can find weaknesses and develop effective quantum error correction or mitigation approaches to improve their performance in real-world, noisy quantum systems. This assures that the algorithms are not only theoretically correct but also practically feasible. These contributions open the way for more resilient and efficient QKD systems, closing the gap between theoretical advances and actual applications in quantum-secure communications. Future research will investigate the suggested algorithms' scalability and adaptability in different and dynamic quantum communication contexts. Overall, integrating QML algorithms, particularly QNN, into QKD protocols has transformative potential since it improves the accuracy, security, and feasibility of key generation processes. These findings highlight the need to use sophisticated quantum learning approaches to address the limitations of traditional and QRL-based quantum key distribution systems, opening the way for next-generation quantum-secure communication frameworks. Future research could focus on optimizing the proposed algorithms and investigating more QML techniques to develop and scale QKD systems for a variety of practical applications.

\ifCLASSOPTIONcaptionsoff
\newpage
\fi

\bibliographystyle{IEEEtran}

\bibliography{IEEE}

\vfill

\end{document}